\documentclass[twocolumn]{aastex62}
\pdfoutput=1 

\usepackage{amsmath}
\usepackage[T1]{fontenc}
\usepackage{url}
\usepackage{graphicx}
\usepackage[caption=false]{subfig}
\usepackage{xcolor}

\received{XXX}
\revised{YYY}
\accepted{ZZZ}

\submitjournal{ApJ}

\shorttitle{Age of Galactic B-type stars}
\shortauthors{Daszy\'nska-Daszkiewicz\& Miszuda}

\begin{document}
	
\title{Age determination of galactic B-type stars in double-lined eclipsing binaries}

\correspondingauthor{Jadwiga Daszy\'nska-Daszkiewicz}
\email{daszynska@astro.uni.wroc.pl}

\author[0000-0001-9704-6408]{Jadwiga Daszy\'nska-Daszkiewicz}
\author{Amadeusz Miszuda}
\affiliation{Astronomical Institute, University of Wroc{\l}aw \\
	ul. Kopernika 11, 51-622 Wroc{\l}aw, Poland}
	
\begin{abstract}
    We present the results of age determination for galactic B-type main sequence stars which are components
    of double-lined eclipsing binaries. Only detached systems are considered.
    We analyze 38 binary systems that meet such criteria.
    The analysis is based on evolutionary computations and we consider that the age is determined if there is
    a common value from the radius$-$age diagrams and the agreement in the position of both components
    in the Hertzsprung-Russell diagram.	In some cases, to meet these two conditions, it was necessary to adjust the value of the metallicity, $Z$
    or/and the parameter of overshooting from the convective core, $\alpha_{\rm ov}$.
    We determine the consistent age for 33 binaries out of 38. Besides, we made an extensive computations
    and for each system we give the range of $\alpha_{\rm ov}$ and $Z$, for which the consistent solutions exist.
    The age of the studied B-type main sequence stars ranges, as counted from the Zero Age Main Sequence,
    from about 2.5 Myr to about 200 Myr.	
\end{abstract}


\section{Introduction} \label{sec:intro}

The knowledge of star's ages is important for many areas of modern astrophysics. For example, the studies of the structure and dynamic
of our Galaxy or formation and evolution of planetary systems require precise stellar ages.
There are several technics for age-dating of stars and ensembles of stars, but no direct method exists.
For stars other than the Sun, usually model-dependent methods are applied.
Till now, one of most accurate methods is based on the mass-radius relation applied to double-lined eclipsing binaries.
The most recent and extensive discussion on age determination and its importance was given by \cite{Soderblom2010} with a special emphasis on low-mass stars.

Mass is the most primary parameter of a star and all evolutionary computations start from fixing its value.
The direct determination of a mass is possible only for binary systems, so-called dynamical mass.
Spectroscopic data alone yield masses with an accuracy of a factor $\sin i$, where $i$ is the inclination angle of the orbital axis.
If the inclination of the orbit is unknown, only a mass ratio can be derived.
The absolute values of masses and radii can be obtained only for double-lined eclipsing binaries.
If eclipses do not occur then long-baseline interferometry may help but such observations are rather rarely available.
Then, for the determination of the absolute values of masses and radii we need the astrometric orbit, angular diameters and accurate parallaxes.

The most simple configurations of a two-star system are detached binaries where individual stellar component's evolution is not altered by the mass transfer.
Usually the coevality and similar chemical compositions for both components are assumed
because they should be formed almost simultaneously from the same cloud of interstellar gas.

The works by \cite{Torres2010} and \cite{Eker2014} constitute excellent compilations of detached double-lined eclipsing binaries (DDLEBs)
with the values of masses and radii determined with an accuracy of 3\% or better.
Because of precise estimates of stellar parameters, double-lined eclipsing binaries
are used as benchmarks for testing the theory of stellar evolution models at various ages and masses.
The  method has been explored since many years. Most recently, \cite{Higl2017} selected 19 systems
with masses spanning from about 0.6 to 14 $M_{\sun}$ to constrain various
parameters of models and theory.  They concluded that in the case of stars less massive than about 1.2 $M_{\sun}$,
there is a need for diffusion. For stars with convective core, they obtained some preference for
overshooting.  An overview of the earlier studies on this subject can be found also in \cite{Higl2017}.

Independent constraints on stellar parameters can be obtained  from asteroseismic modelling.
If pulsational modes can be well identified then, both, parameters of a model (e.g., mass, effective temperature, radius) and parameters of theory
(e.g., overshooting form convective regions, opacity data) can be constrained .
Thus, double-lined eclipsing binary with pulsating components suitable for seismic studies would be the best gauges,
in particular, for calibration of free parameters that cannot be derived from first principles, e.g.,
the amount of convective overshooting, mixing length parameter describing the efficiency of convective transport
or parameters of rotational-induced mixing.

Up to now there is no in-depth studies of such case for massive stars because there are not many double-lined eclipsing massive binaries
with pulsating components or there is a lack of reliable identification of pulsational modes. The promising case is a single-lined eclipsing
binary V381 Car with a pulsating primary \citep{Jerzykiewicz1992, Freyhammer2005}. Till now three frequencies were detected corresponding
to low order p modes \citep{Freyhammer2006}. However, the simultaneous binary and asteroseismic modelling of the V381 Car system awaits
more time series photometric and spectroscopic data in order to detect spectral lines of the secondary star and to get unambiguous identification of pulsational modes.

Recently, an attempt to such combined analysis was made by \cite{Tkachenko2016} for Spica which is a spectroscopic binary with pulsating component
but not the eclipsing one. Unfortunately, insufficient frequency resolution and detection of only three pulsational frequencies with no firm mode identification
prevented detailed seismic modelling.

Here, we present the determination of age for a homogenous sample of stars. We focus on all double-lined eclipsing binaries with two B-type components
on main sequence for which accurate values (i.e., usually below 3\%) of masses and radii were derived. Based on works by \cite{Torres2010} and \cite{Eker2014},
this sample consists of 38 binary systems. The more recent determination of masses and radii were also included in some cases.
We chose B-type stars because of their importance for evolution of the chemical composition and the structure of galaxies.
Moreover, stars with masses above 8$M_{\sun}$ produce a collapsing core and explode as a supernova.
Therefore, knowledge of properties of B-type main-sequence stars and their age is crucial for a better understanding of next evolutionary stages.
An important argument is also that in comparison to more or less massive stars, their structure and related phenomena are relatively simple. For example
we do not encounter such problems as strong mass loss or efficient convection in the outer layers. Thus, B-type eclipsing systems of the SB2 type
are also good indicators of age for open cluster and associations.
In this paper we apply the definition of the zero age as a time on the Zero Age Main Sequence (ZAMS).

In Sect.\,2, we introduce the selected sample of B-type binary stars. Sect.\,3 is devoted to the assessment of uncertainties in estimation of age
resulting from the adopted theoretical parameters describing various effects, i.e., metallicity, initial hydrogen abundance, rotation and overshooting from the convective core.
Results of the age determinations are given in Sect.\,4 and a few examples are described in details. Conclusions and future plans close the paper.

\section{The sample of binary stars}

\begin{figure}
	\includegraphics[width=\columnwidth, clip]{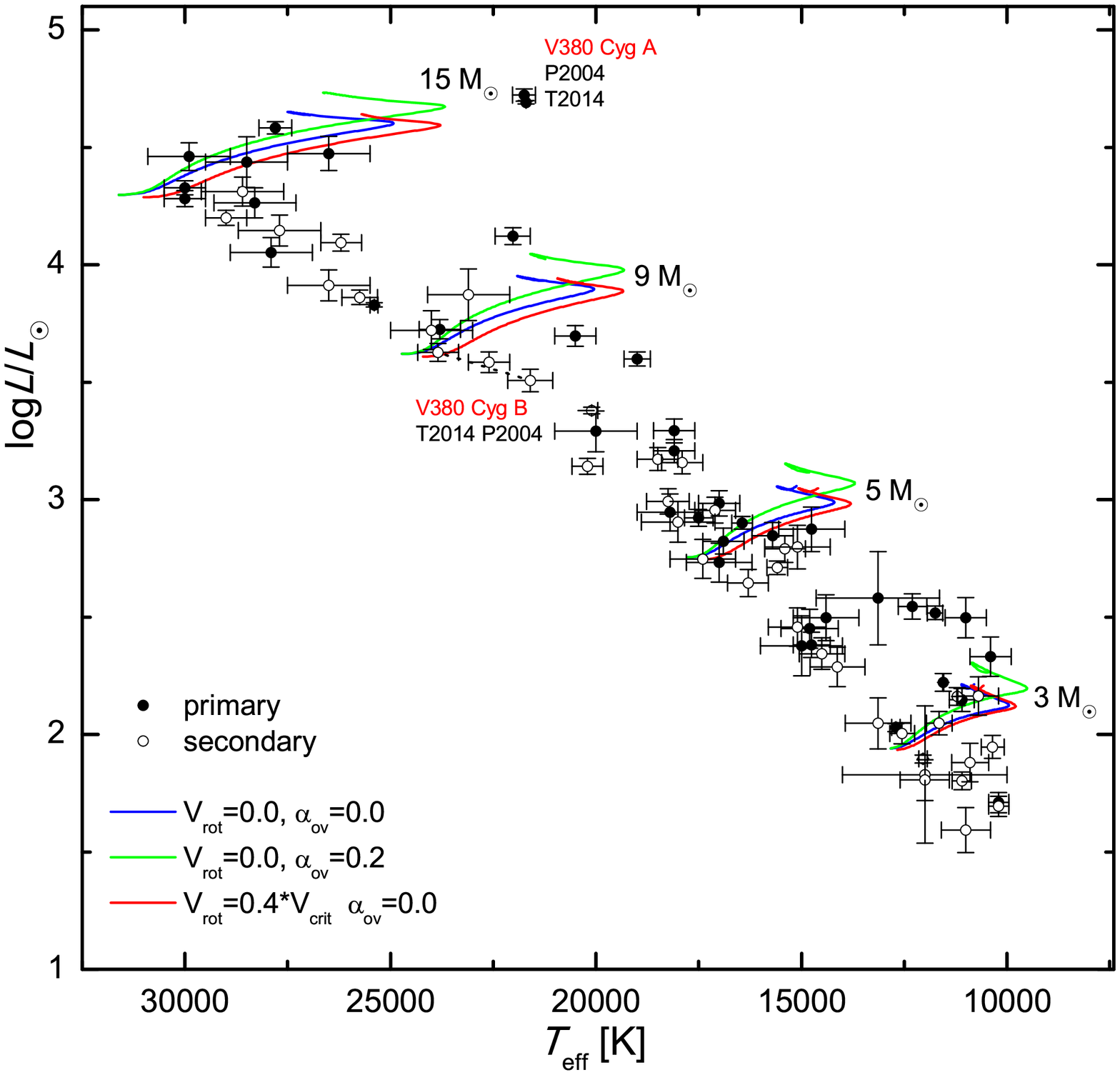}
	\caption{The position of components of our sample of 38 binaries on the Hertzsprung-Russell diagram.
 The evolutionary tracks were computed for the initial hydrogen abundance $X_0=0.70$, metallicity $Z=0.014$ and the AGSS09 solar mixture. 
There is shown the effect of rotation, $V_{\rm rot}$, and overshooting from the convective core, $\alpha_{\rm ov}$.} 
	\label{fig1}
\end{figure}
\begin{figure}
	\includegraphics[width=\columnwidth, clip]{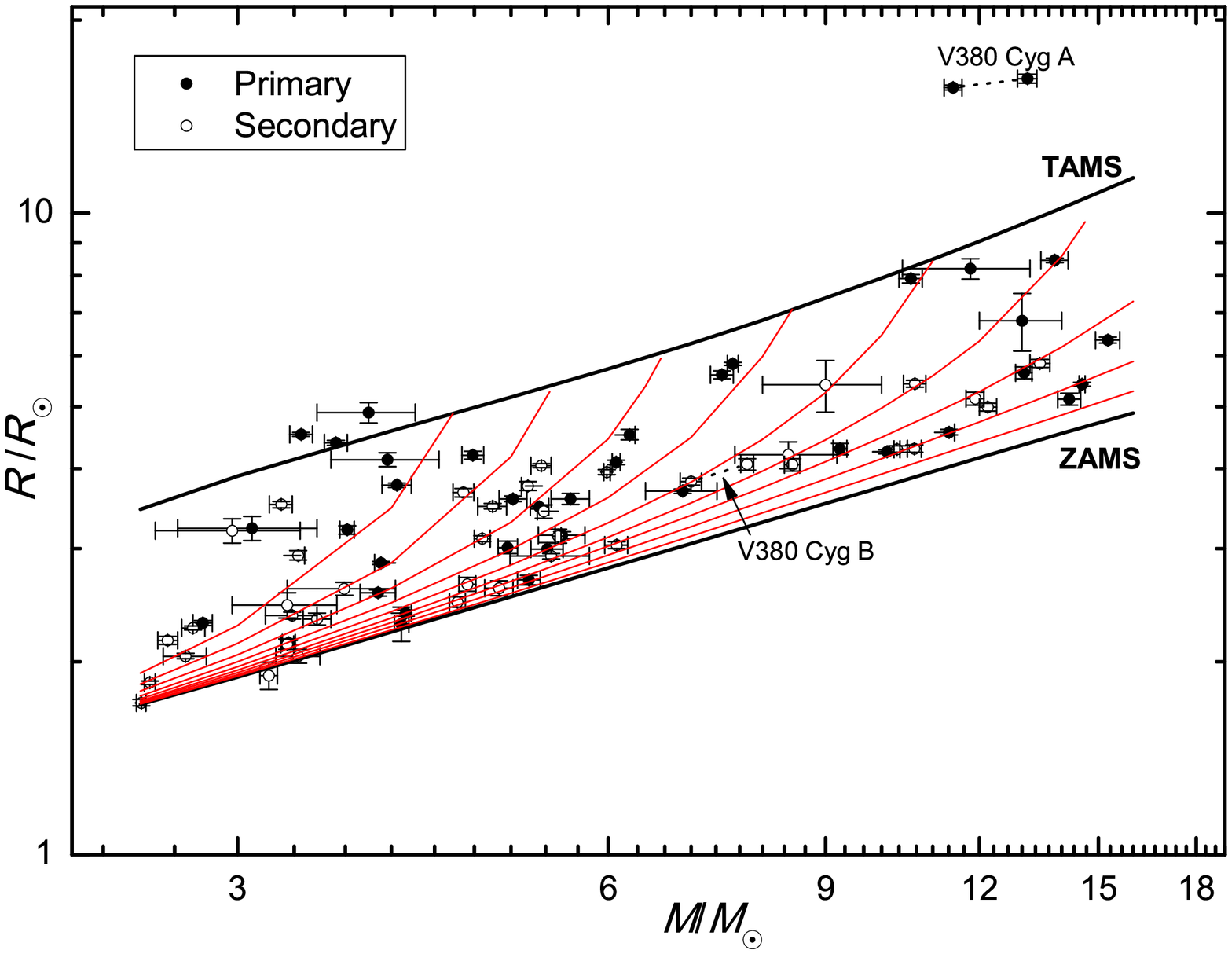}
	\caption{The mass-radius diagram with the positions of the components of 38 selected binaries. There are shown also the lines of
		ZAMS, TAMS} and isochrones from $\log (t/{\rm yr})=6$ to $\log (t/{\rm yr})=8$.
		They were computed for the metallicity $Z=0.014$, initial hydrogen abundance $X_0=0.70$, without including rotation and convective core overshooting.
	\label{fig2}
\end{figure}
\begin{figure}
	\includegraphics[width=\columnwidth, clip]{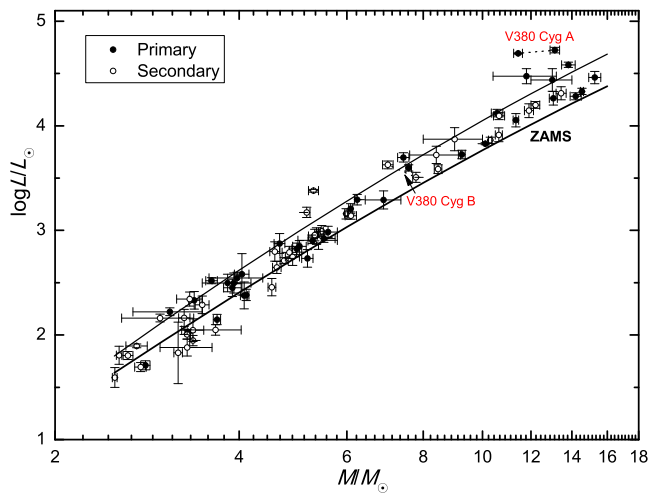}
	\caption{The mass-luminosity diagram with the positions of components of 38 selected binaries. The lines of ZAMS and TAMS are for the same parameters as in Fig.\,2.}
	\label{fig3}
\end{figure}

For age determination we selected uniform sample of binary stars.
All of them are double-lined eclipsing binaries with two components of B spectral type evolving on main sequence.
All of them are considered to be detached binary systems thus, in the first approximation, one can assume that there is no interaction between both components and their evolution
proceeds independently.
We use the data on dynamical masses and radii from \cite{Torres2010} supplemented with the more recent catalogue
by \cite{Eker2014} as well as other recent determinations. These parameters are known in most cases with an accuracy of 3\% or better

In total, there are 38 DDLEBs that meet the above criteria. In Table\,1, we list the basic parameters of these stars, i.e.,
the name, HD number, spectral types, the orbital period, $P$, the maximum brightness in the Johnson $V$ filter, $V_{\rm max}$,
mass, $M$, radius, $R$,  effective temperature, $T_{\rm{eff}}$, luminosity, $\log{L/L_{\odot}}$, metallicity [m/H],
the projected value of the rotational velocity, $V_{\rm rot}\sin i$, the approximate value of the synchronous rotation.
In case of the systems AH  Cep and V380 Cyg, two sets of stellar parameters were listed, from two different sources.
In Fig.\,1 we depicted their positions on the Hertzsprung-Russell diagram. The values of the effective temperature, $T_{\rm eff}$,
were taken from \cite{Torres2010} and \cite{Eker2014}. The values of luminosities, $L/L_{\sun}$, were computed from
the effective temperatures, $T_{\rm eff}$, and radii, $R$.
In the case of V380 Cyg, the parameters from two determinations were included, i.e., from Pavlovski \& Southworth (2009) and Tkachenko et al. (2014).
The evolutionary tracks for four values of masses (3, 5, 9, 15 $M_{\sun}$) were depicted.
They were computed with the Warsaw$-$New Jersey code \citep[e.g.,][]{Pamyatnykh1999} assuming the initial hydrogen abundance by mass $X_0=0.70$,
metallicity Z=0.014, OPAL opacities \citep{Iglesias1996} and the solar heavy element mixture by \cite{Asplund2009}, hereafter AGSS09.
The effects of rotation and overshooting from the convective core on the evolutionary tracks are shown as well.
In this paper, we adopted the abundance of metallicity by mass $Z=0.014$ as determined by \cite{Nieva2012} for galactic B-type stars.

From Figs.\,2 and 3 one can see that these stars obey quite well  the  mass-radius and mass-luminosity relations, respectively.
One star clearly stands out from these dependencies. This is the case of the more massive component of the binary V380 Cyg, whose
evolutionary stage is still uncertain. We show the two sets of parameters of V380 Cyg,
as determined by \cite{Pavlovski2009} and \cite{Tkachenko2014}, because they are quite different.

In the mass-radius diagram shown in Fig.\,2, we plotted the lines of ZAMS (Zero Age Main Sequence) and TAMS (Terminal Age Main Sequence) as well as
isochrones from the age $\log (t/{\rm yr})=6.0$ to $\log (t/{\rm yr})=8.0$, assuming the initial hydrogen abundance $X_0=0.70$, metallicity $Z=0.014$, no-rotation and no-overshooting.
Such diagrams are often used for the age determination from accurate masses and radii.

The values of the basic orbital parameters of the selected binary systems are depicted as histograms in Fig.\,4.
As one can see, most of the systems have the mass ratio, $M_B/M_A$, higher than 0.7 and only three systems have $M_B/M_A<0.6$; these are V380 Cyg, V379 Cep and V1331 Aql.
The majority of the systems (27) have short orbital periods with $P_{\rm orb}< 5$\,d and their orbits are not very elongated, i.e.,
$e<0.3$ for 34 systems. The system with the longest orbital period, $P\approx 100$ \,d, is V379 Cep which is a quadruple system with two binaries \citep{Harmanec2007}.
One of them is the eclipsing binary consisting of two B-type stars. The inclination angle of the orbital axis is larger than $i=75^\circ$ for 31 binaries.
Two systems have the value of $i$ below $65^\circ$, these are IM Mon \citep{Bakis2011} and V497 Cep \citep{Yakut2007}.

\startlongtable
\begin{deluxetable*}{lclcccccclll}
	\tablecaption{Parameters of double-lined eclipsing binaries with two components of B spectral type.
	 Detailed description is given in the text. The last column contains the references.\label{chartable}}
	\tabletypesize{\scriptsize}
	\tablehead{
		\colhead{System} & \colhead{Star} &
		\colhead{SpT} & \colhead{$P$} & \colhead{$M$} &
		\colhead{$R$} & \colhead{$T_{\rm eff}$} &
		\colhead{$\log L/L_{\odot}$ } & \colhead{[m/H]} &
		\colhead{$V_{\rm rot} \sin i$ } & \colhead{$V_{\rm synch}$} & \colhead{Ref.} \\
		\colhead{} & \colhead{} & \colhead{} & \colhead{$V_{\rm max}$} & \colhead{[$M_{\odot}$]} &
		\colhead{[$R_{\odot}$]} & \colhead{[K]} & \colhead{} &
		\colhead{} & \colhead{[km/s]} & \colhead{[km/s]} & \colhead{}
	}
	\startdata
		AH Cep			& A	& B0.5Vn		& 1.77		& 15.26$\pm$0.35				& 6.346$\pm$0.071			& 29900$\pm$1000				& 4.461$\pm$0.059				& --										& 185$\pm$30 		& 181			&	1		\\
		HD 216014	& B	& B0.5Vn		& 6.81		& 13.44$\pm$0.25				& 5.836$\pm$0.085			& 28600$\pm$1000				& 4.311$\pm$0.062				& 											&  185$\pm$30 		& 167			&		 \vspace{5pt}\\
					& A	& B0.2V		& 1.77		& 14.30$\pm$1.00				& 5.60$\pm$0.10			& 31000$\pm$3000				& 4.415$\pm$0.169				& --										& 200 						& 160			&	13		\\
		        	& B	& B2V			& 6.88		& 12.60$\pm$0.90				& 4.70$\pm$0.10			& 29000$\pm$4000				& 4.147$\pm$0.240				& 											& 170 						& 134			&		 \vspace{5pt}\\
		V578 Mon	& A	& B1V			& 2.41		& 14.54$\pm$0.08				& 5.41$\pm$0.04			& 30000$\pm$500					& 4.330$\pm$0.030				& -0.30$\pm$0.13				& 117$\pm$5 		& 109			&	1,14,15		\\
		HD 259135	& B	& B2V			& 8.55		& 10.29$\pm$0.06				& 4.29$\pm$0.05			& 25750$\pm$435					& 3.860$\pm$0.031				& 											&  94$\pm$4 			& 90				&  	\vspace{5pt}\\
		HI Mon 			& A	& B0V			& 1.57		& 14.20$\pm$0.30  			& 5.130$\pm$0.110			& 30000$\pm$500					& 4.282$\pm$0.034				& --										& 150$\pm$25 		& 165			& 	2		\\
		HD 51076	& B	& B0.5V		& 9.45		& 12.20$\pm$0.20				& 4.990$\pm$0.070			& 29000$\pm$500 				& 4.199$\pm$0.032				& 											& 150$\pm$25 		& 160			&		\vspace{5pt}\\
		V453 Cyg 	& A	& B0.4IV		& 3.89		& 13.82$\pm$0.35				& 8.445$\pm$0.068			& 27800$\pm$400					& 4.583$\pm$0.026				& -0.25$\pm$0.05				& 109$\pm$3			& 111			&  1,16,17		\\
		HD 227696	& B	& B0.7IV		& 8.28		& 10.64$\pm$0.22				& 5.420$\pm$0.068			& 26200$\pm$500					& 4.094$\pm$0.035				&											& 98$\pm$5 			& 71				& 		\vspace{5pt}\\
		V380 Cyg		& A	& B1.5II-III	& 12.43	& 13.13$\pm$0.24  			& 16.22$\pm$0.26			& 21750$\pm$280					& 4.723$\pm$0.026				& 0.05$\pm$0.12				& 98$\pm$4 			& 66				&	3, 7		\\
		HD 187879	& B	& B2V			& 5.68		& 7.78$\pm$0.10				& 4.060$\pm$0.084			& 21600$\pm$550					& 3.508$\pm$0.048				& 	 										& 32$\pm$6 			& 17				&		\vspace{5pt}\\
		    		& A	& B1.5II-III	& 12.43	& 11.43$\pm$0.19  			& 15.71$\pm$0.13			& 21700(fixed)						& 4.691$\pm$0.007				&    				& 98$\pm$4 			& 64				&	4, 7		\\
		        	& B	& B2V			& 5.68		& 7.00$\pm$0.14				& 3.819$\pm$0.048			& 23840$\pm$500					& 3.626$\pm$0.038				& 	 										& 32$\pm$6 			& 16				&		\vspace{5pt}\\
		CW Cep		& A	& B0.5V		& 2.73		& 13.05$\pm$0.20				& 5.640$\pm$0.120			& 28300$\pm$1000				& 4.263$\pm$0.064				& --										& 132  					& 102			&	1, 6		\\
		HD 218066	& B	& B0.5V		& 	7.59		& 11.91$\pm$0.20				& 5.140$\pm$0.120			& 27700$\pm$1000				& 4.145$\pm$0.066				& 											& 138 						& 93				&		\vspace{5pt}\\
		NY Cep			& A	& B0.5V		& 15.28	& 13.0$\pm$1.0 			& 6.800$\pm$0.700			& 28500$\pm$1000				& 4.438$\pm$0.108				& --										& 75$\pm$10 		& 23				&	2		\\
		HD 217312	& B	& B2V			& 	7.43		& 9.0$\pm$1.0				& 5.400$\pm$0.500			& 23100$\pm$1000				& 3.872$\pm$0.110				&  										& 125$\pm$14 		& 18				&		\vspace{5pt}\\
		V346 Cen		& A	& B1.5III		& 6.32		& 11.8$\pm$1.4  			& 8.2$\pm$0.3 		& 26500$\pm$1000					& 4.474$\pm$0.073				& --										& 165$\pm$15				 		& 66				&	2, 9 \\
		HD 101837	& B	& B2V			& 8.54		& 8.4$\pm$0.8				& 4.2$\pm$0.2			& 24000$\pm$1000					& 3.721$\pm$0.083				&  										& 140$\pm$15		 		& 34				&		\vspace{5pt}\\
		DW Car 		& A	& B1V			& 1.33		& 11.34$\pm$0.18				& 4.561$\pm$0.050			& 27900$\pm$1000				& 4.054$\pm$0.063				& --										& 182$\pm$3  		& 174			&	1		\\
		HD 305543	& B	& B1V			& 	9.68		& 10.63$\pm$0.20				& 4.299$\pm$0.058			& 26500$\pm$1000				& 3.913$\pm$0.066				& 											& 177$\pm$3 		& 164			&		\vspace{5pt}\\
		V379 Cep		& A	& B2IV			& 99.76	& 10.56$\pm$0.23  			& 7.909$\pm$0.120			& 22025$\pm$428					& 4.121$\pm$0.036				& --										& -- 							& 4				&	2		\\
		HD 197770	& B	& ---				& 6.33		& 6.09$\pm$0.13				& 3.040$\pm$0.040			& 20206$\pm$374					& 3.141$\pm$0.034				&  										& --							& 2				&		\vspace{5pt}\\
		V1331 Aql	& A	& B1V			& 1.36		& 10.10$\pm$0.11  			& 4.240$\pm$0.030			& 25400$\pm$100					& 3.829$\pm$0.009				& --										& -- 							& 158			&	2		\\
		HD 173198	& B	& B1.5V		& 7.80		& 5.29$\pm$0.10				& 4.040$\pm$0.030			& 20100$\pm$140					& 3.379$\pm$0.014				&  										& --							& 150			& 		\vspace{5pt}\\
		QX Car 			& A 	& B2V 			& 4.48 	& 9.25$\pm$0.12 					& 4.291$\pm$0.091 			& 23800$\pm$500 				& 3.725$\pm$0.041 				& -- 										& 120$\pm$10  		& 48				&	1		\\
		HD 86118 	& B 	& B2V 			& 6.64		& 8.46$\pm$0.12 					& 4.053$\pm$0.091 			& 22600$\pm$500 				& 3.585$\pm$0.043 				&  										& 110$\pm$10 		& 46				&		\vspace{5pt}\\
		V399 Vul		& A 	& B3IV-V 		& 4.90 	& 7.57$\pm$0.08 				& 5.820$\pm$0.030 			& 19000$\pm$320 				& 3.598$\pm$0.030				& -- 										& 61$\pm$5  			& 60				&	2		\\
		HD 194495 	& B 	& B4V 			& 7.06		& 5.46$\pm$0.03 				& 3.140$\pm$0.080 			& 18250$\pm$520 				& 2.992$\pm$0.054				&  										& 39$\pm$7 			& 32				&		\vspace{5pt}\\
		V1388 Ori 	& A 	& B2.5IV-V 	& 2.19 	& 7.42$\pm$0.16 				& 5.600$\pm$0.080			& 20500$\pm$500 				& 3.697$\pm$0.044 				& -- 										& 125$\pm$10  		& 130			&	1		\\
		HD 42401 	& B 	& B3V 			& 7.40		& 5.16$\pm$0.06 				& 3.760$\pm$0.060			& 18500$\pm$500 				& 3.172$\pm$0.049 				&  										& 75$\pm$15 		& 87				&		\vspace{5pt}\\
		V497 Cep 	& A 	& B3V 			& 1.20		& 6.89$\pm$0.46 				& 3.690$\pm$0.030 			& 19500$\pm$950 				& 3.245$\pm$0.087 				& -- 										& -- 							& 155			&	5		\\
		BD+61 2213& B 	& B4V 			& 8.95		& 5.39$\pm$0.40 				& 2.920$\pm$0.030			& 17500$\pm$950 				& 2.877$\pm$0.087 				&  										& -- 							& 123			&		\vspace{5pt}\\
		V539 Ara 	& A 	& B3V 			& 3.17 	& 6.240$\pm$0.066 				& 4.516$\pm$0.084 			& 18100$\pm$500 				& 3.293$\pm$0.051 				& -- 										& 75$\pm$8  			& 71				&	1		\\
		HD 161783 	& B 	& B4V 			& 5.71		& 5.314$\pm$0.060 				& 3.428$\pm$0.083 			& 17100$\pm$500 				& 2.955$\pm$0.055 				&  										& 48$\pm$5 			& 60				&		\vspace{5pt}\\
		CV Vel 			& A 	& B2.5V 		& 6.89 	& 6.086$\pm$0.044 				& 4.089$\pm$0.036 			& 18100$\pm$500 				& 3.207$\pm$0.049 				&  --       & 31$\pm$2			& 30				&	1		\\
		HD 77464 	& B 	& B2.5V 		& 6.69		& 5.982$\pm$0.035 				& 3.950$\pm$0.036 			& 17900$\pm$500 				& 3.158$\pm$0.049 				&  	 									& 19$\pm$1 			& 29				&		\vspace{5pt}\\
		LT CMa			& A 	& B4V 			& 1.76		& 5.59$\pm$0.20 				& 3.590$\pm$0.070 			& 17000$\pm$500 				& 2.985$\pm$0.054				& -- 										& 109$\pm$10  		& 103			&	2		\\
		HD 53303 	& B 	& B6.5V 		& 7.44		& 3.36$\pm$0.14 				& 2.040$\pm$0.050 			& 13140$\pm$800 				& 2.047$\pm$0.108				&  										& 67$\pm$10 		& 59				&		\vspace{5pt}\\
		IM Mon			& A 	& B4V 			& 1.19 	& 5.50$\pm$0.24 				& 3.150$\pm$0.040 			& 17500$\pm$350 				& 2.922$\pm$0.036				& 0.2$\pm$0.15 				& 147$\pm$15  		& 134			&	2,12		\\
		HD 44701 	& B 	& B6.5V		& 6.55		& 3.32$\pm$0.16 				& 2.360$\pm$0.030 			& 14500$\pm$550 				& 2.344$\pm$0.067				&  										& 90$\pm$25 		& 100			&		\vspace{5pt}\\
		AG Per 			& A 	& B3.4V 		& 2.03 	& 5.359$\pm$0.160 				& 2.995$\pm$0.071 			& 18200$\pm$800 				& 2.946$\pm$0.079 				& -- 										& 94$\pm$23  		& 75				&	1		\\
		HD 25833 	& B 	& B3.5V 		& 6.72		& 4.890$\pm$0.130 				& 2.605$\pm$0.070 			& 17400$\pm$800 				& 2.747$\pm$0.083 				&  										& 70$\pm$9 			& 65				&		\vspace{5pt}\\
		U Oph 			& A 	& B5V 			& 1.68 	& 5.273$\pm$0.091 				& 3.484$\pm$0.021 			& 16440$\pm$250 				& 2.901$\pm$0.027 				& -- 										& 125$\pm$5 		& 100			&	1		\\
		HD 156247 	& B 	& B6V 			& 5.87		& 4.739$\pm$0.072 				& 3.110$\pm$0.034 			& 15590$\pm$250 				& 2.710$\pm$0.029 				&  										& 115$\pm$5 		& 91				&		\vspace{5pt}\\
		DI Her 			& A 	& B5V 			& 10.55 	& 5.170$\pm$0.110 				& 2.681$\pm$0.046 			& 17000$\pm$800 				& 2.732$\pm$0.083 				& -- 										& 108 						& 13				&	1, 8		\\
		HD 175227 	& B 	& B5V 			& 8.42		& 4.524$\pm$0.066 				& 2.478$\pm$0.046 			& 15100$\pm$700 				& 2.457$\pm$0.082 				&  										& 116 						& 12				&		\vspace{5pt}\\
		EP Cru			& A 	& B5V 			& 11.08	& 5.020$\pm$0.130 				& 3.590$\pm$0.035 			& 15700$\pm$500 				& 2.847$\pm$0.056				& -- 										& 141$\pm$5  		& 16				&	2		\\
		HD 109724	& B 	& B5V			& 8.69		& 4.830$\pm$0.130 				& 3.495$\pm$0.034 			& 15400$\pm$500 				& 2.790$\pm$0.057				&  										& 138$\pm$5			& 16				&		\vspace{5pt}\\
		V760 Sco 	& A 	& B4V 			& 1.73 	& 4.969$\pm$0.090 				& 3.015$\pm$0.066 			& 16900$\pm$500 				& 2.823$\pm$0.055 				& -- 										& 95$\pm$10			& 88				&	1		\\
		HD 147683 	& B 	& B4V 			& 6.99		& 4.609$\pm$0.073 				& 2.641$\pm$0.066 			& 16300$\pm$500 				& 2.645$\pm$0.058 				&  										& 85$\pm$10 		& 77				&		\vspace{5pt}\\
		MU Cas 		& A 	& B5V 			& 9.65 	& 4.657$\pm$0.093 				& 4.195$\pm$0.058 			& 14750$\pm$800 				& 2.874$\pm$0.095 				& 0.22									& 21$\pm$2 			& 22				&	1, 18		\\
		BD+59 22	& B 	& B5V 			& 10.80	& 4.575$\pm$0.088 				& 3.670$\pm$0.057 			& 15100$\pm$800 				& 2.798$\pm$0.093 				&  										& 22$\pm$2 			& 19				&		\vspace{5pt}\\
		GG Lup 		& A 	& B7V 			& 1.85 	& 4.106$\pm$0.044 				& 2.380$\pm$0.025 			& 14750$\pm$450 				& 2.382$\pm$0.054 				& -- 										& 97$\pm$8 			& 65				&	1		\\
		HD 135876 	& B 	& B9V 			& 5.59		& 2.504$\pm$0.023 				& 1.726$\pm$0.019 			& 11000$\pm$600 				& 1.593$\pm$0.095 				&  										& 61$\pm$5 			& 47				&		\vspace{5pt}\\
		V615 Per 	& A 	& B7V 			& 13.72 	& 4.075$\pm$0.055 				& 2.291$\pm$0.141 			& 15000$\pm$1100 				& 2.370$\pm$0.128 				& $-0.3$ 					& 28$\pm$5 			& 8				&	5, 11		\\
		& B 	& ---				& 13.02	& 3.179$\pm$0.051 				& 1.903$\pm$0.094 			& 13000$\pm$1300 				& 1.960$\pm$0.293 				&  										& 8$\pm$5 				& 7				&		\vspace{5pt}\\
		BD+03\_3821& A & B8V 		& 3.66		& 4.040$\pm$0.110 				& 3.770$\pm$0.030 			& 13140$\pm$1500 				& 2.580$\pm$0.198				& -- 										& 109$\pm$2  		& 52				&	2		\\
		HD 174884	& B 	& ---				& 7.98		& 2.720$\pm$0.110 				& 2.040$\pm$0.020 			& 12044$\pm$100 				& 1.896$\pm$0.017				&  										& 60$\pm$3 			& 28				&		\vspace{5pt}\\
		V1665 Aql	& A 	& B9V 			& 3.88		& 3.970$\pm$0.400 				& 4.130$\pm$0.100 			& 12300$\pm$350 				& 2.545$\pm$0.054				& -- 										& --  						& 54				&	2		\\
		HD 175677	& B 	& ---				& 8.11		& 3.660$\pm$0.370 				& 2.600$\pm$0.060 			& 11650$\pm$310 				& 2.048$\pm$0.050				&  										& -- 							& 34				&		\vspace{5pt}\\
		$\zeta$ Phe	& A 	& B6V 			& 1.67 	& 3.921$\pm$0.045 				& 2.852$\pm$0.015 			& 14400$\pm$800 				& 2.497$\pm$0.097 				& -- 										& 85$\pm$8 			& 86				&	1		\\
		HD 6882 		& B 	& B8V 			& 3.95		& 2.545$\pm$0.026 				& 1.854$\pm$0.011 			& 12000$\pm$600 				& 1.806$\pm$0.087 				&  										& 75$\pm$8 			& 56				&		\vspace{5pt}\\
		YY Sgr			& A 	& B5V 			& 2.63		& 3.900$\pm$0.130 				& 2.560$\pm$0.030 			& 14800$\pm$700 				& 2.451$\pm$0.083				& -- 										& 58$\pm$2  			& 49				&	2, 10		\\
		HD 173140 	& B 	& B6V			& 10.17	& 3.480$\pm$0.090 				& 2.330$\pm$0.050 			& 14125$\pm$670 				& 2.288$\pm$0.084				&  										& 39$\pm$2 			& 45				&		\vspace{5pt}\\
		V398 Lac		& A 	& B9V 			& 5.41		& 3.830$\pm$0.350 				& 4.890$\pm$0.180 			& 11000$\pm$500 				& 2.497$\pm$0.085				& -- 										& 79$\pm$2  			& 46				&	2		\\
		HD 210180	& B 	& ---				& 8.75		& 3.290$\pm$0.320 				& 2.450$\pm$0.110 			& 10900$\pm$450 				& 1.881$\pm$0.082				&  										& 19$\pm$2 			& 23				&		\vspace{5pt}\\
		V413 Ser		& A 	& B8V 			& 2.26		& 3.680$\pm$0.050 				& 3.210$\pm$0.050 			& 11100$\pm$300 				& 2.147$\pm$0.049				& -- 										& 44$\pm$5  			& 72				&	2,10		\\
		HD 171491 	& B 	& B9V			& 7.99		& 3.360$\pm$0.040 				& 2.930$\pm$0.050 			& 10350$\pm$280 				& 1.947$\pm$0.049				&  										& 38$\pm$7 			& 66				&		\vspace{5pt}\\
		$\chi^2$ Hya& A	& B8V 			& 2.27 	& 3.605$\pm$0.078 				& 4.390$\pm$0.039 			& 11750$\pm$190 				& 2.518$\pm$0.029 				& -- 										& 112$\pm$10 		& 98				&	1		\\
		HD 96314 	& B 	& B8V 			& 5.65		& 2.632$\pm$0.049 				& 2.159$\pm$0.030 			& 11100$\pm$230 				& 1.803$\pm$0.038 				&  										& 60$\pm$6 			& 48				&		\vspace{5pt}\\
		V906 Sco 	& A 	& B9V 			& 2.79 	& 3.378$\pm$0.071 				& 4.521$\pm$0.035 			& 10400$\pm$500 				& 2.332$\pm$0.084 				& 0.14$\pm$0.06 				& 80$\pm$5 			& 82				&	1,19,20		\\
		HD 162724 	& B 	& B9V 			& 5.96		& 3.253$\pm$0.069 				& 3.515$\pm$0.039 			& 10700$\pm$500 				& 2.163$\pm$0.082 				& 0.03$\pm$0.02				& 62$\pm$8 			& 64				&		\vspace{5pt}\\
		$\eta$ Mus	& A 	& B8V 			& 2.40		& 3.300$\pm$0.040 				& 2.140$\pm$0.020 			& 12700$\pm$100 				& 2.029$\pm$0.016				& -- 										& 34$\pm$2  			& 45				&	2		\\
		HD 114911	& B 	& B8V			& 4.78		& 3.290$\pm$0.040 				& 2.130$\pm$0.040 			& 12550$\pm$300 				& 2.005$\pm$0.045				&  										& 44$\pm$2 			& 45				&		\vspace{5pt}\\
		V799 Cas		& A 	& B8V 			& 7.70		& 3.080$\pm$0.400 				& 3.230$\pm$0.140 			& 11550$\pm$14 					& 2.222$\pm$0.038				& -- 										& --  						& 21				&	2,10		\\
		HD 18915 	& B 	& B8.5V		& 8.82		& 2.970$\pm$0.400 				& 3.200$\pm$0.140 			& 11210$\pm$14 					& 2.162$\pm$0.038				&  										& -- 							& 21				&		\vspace{5pt}\\
		PV Cas 		& A 	& B9.5V 		& 1.75 	& 2.816$\pm$0.050 				& 2.301$\pm$0.020 			& 10200$\pm$250 				& 1.711$\pm$0.043 				& -- 										& -- 							& 66				&	1		\\
		HD 240208 	& B 	& B9.5V 		& 9.72		& 2.757$\pm$0.054 				& 2.257$\pm$0.019 			& 10190$\pm$250 				& 1.693$\pm$0.043 				&  										& --  						& 65				&		\vspace{5pt}\\
\enddata
\tablecomments{1) \cite{Torres2010}, 2) \cite{Eker2014}, 3) \cite{Pavlovski2009},
	4) \cite{Tkachenko2014}, 5) \cite{Yildiz2011}, 6) \cite{Nha1975}, 7) \cite{Prugniel2011}, 8) \cite{Philippov2013},
	9) \cite{Mayer2016}, 10) \cite{Hog2000}, 11) \cite{Gonzales2000}, 12) \cite{Bakis2011}, 13) \cite{Martins2017}, 14) \cite{Pavlovski2005},
	15) \cite{Garcia2014}, 16) \cite{Pavlovski2009A}, 17) \cite{Southworth2004}, 18) \cite{Lacy2004}, 19) \cite{Sestito2003}, 20) \cite{Villanova2009}}
\end{deluxetable*}

\begin{figure*}
	\includegraphics[width=\textwidth,clip]{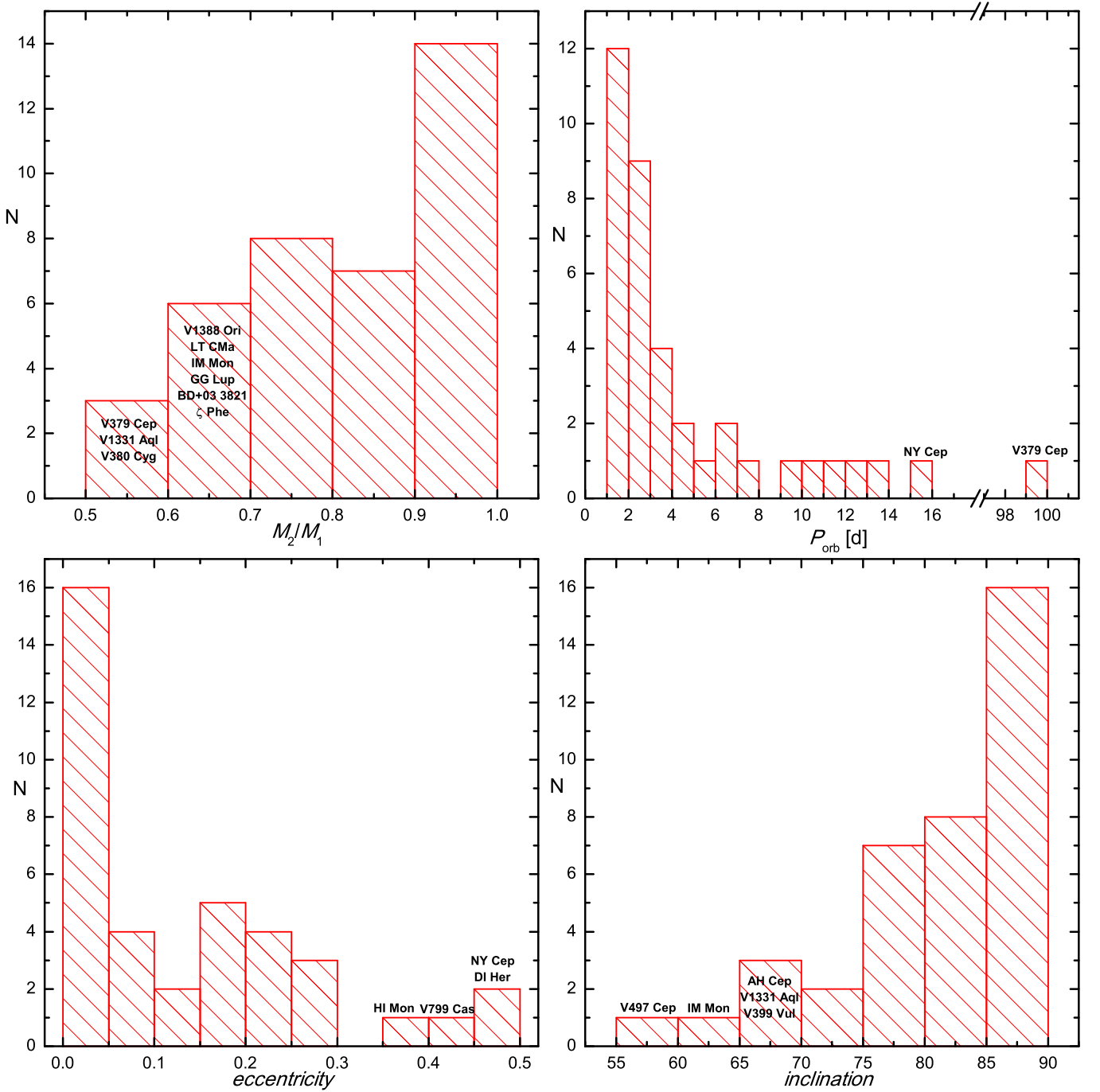}
	\caption{Histograms for main orbital parameters of 38 main sequence eclipsing binaries with B-type components.}
	\label{hist}
\end{figure*}

\section{Age from accurate masses and radii: theoretical uncertainties}

As was demonstrated in the previous section the majority of binaries considered here have short orbital periods and low eccentricities.
This is very likely a manifestation of the tidal interaction between the components that leads to synchronisation and circularisation \citep[e.g.,][]{Siess2013}.
Thus in most cases the interaction can occur but without any signatures of mass transfer.
Therefore, in this paper we assume that, in the first approximation, each star evolves separately and we rely on single-star evolutionary code.
However, at some point the results should be confronted with advanced approach based on binary evolution models.

A grid of evolutionary models was computed using the Warsaw-New Jersey code \citep[e.g.,][]{Pamyatnykh1999}.
We adopted OPAL opacity tables \citep{Iglesias1996}, the initial hydrogen abundance by mass $X_0=0.70$
and the AGSS09 solar mixture of heavy elements.

The Warsaw-New Jersey code includes the effects of solid rotation, in particular the mean effects of centrifugal force,
assuming that global angular momentum is conserved during evolution.
Here, we neglected the effect of mass loss, because even for the hottest main sequence B stars, the mass-loss rate resulting
from line-driven winds is of the order of $10^{-9}~M_{\sun}/{\rm yr}$ \citep[e.g.,][]{Krticka2014}.
Thus, assuming the mean age of early B-type stars on the main sequence of the order of $10^7$\,yr one gets that a star loses
a mass of the order of hundredths of solar mass.

Overshooting from the convective core is included according to \cite{Dziembowski2008}.
This is a two-parameter prescription that allows for non-zero gradient of the hydrogen abundance inside
the partly mixed region above the convective core. The extent of overshooting is measured by $\alpha_{\rm ov}H_p$
where $H_p$ is the pressure height scale and $\alpha_{\rm ov}$ is a free, unknown parameter.

In the considered sample of stars the energy transport by convection in envelope can be neglected.
Also, as has been mentioned, in the considered range of masses ($\sim$3-15$M_\odot$), there is no need to include the effect of mass loss.
Thus, we get rid of the uncertainty associated with two additional free parameters.

The dependency between radius and mass at various ages is the classical diagram used for the age determination.
In Fig.\,2, we plotted the isochrones computed for the standard chemical composition from ZAMS to TAMS.
Only the mass range corresponding to the B-type stars was considered.
However, using the mass-radius (MR) diagram demands a number of interpolations to derive an accurate age.
The radius-age diagrams at the fixed values of masses are much simpler to use \citep[e.g.,][]{Higl2017}.
Any of these approaches  contains uncertainties associated with the values of the parameters used for theoretical calculations.

In Fig.\,5, we show the value of radius, $R$, as a function of age, $\log (t/{\rm yr})$, for the four values
of masses, $M=3$, 5, 9 and 15\,$M_{\sun}$.
There are shown also the effects of various parameters in each panel, namely the effect of metallicity, $Z$, (the left$-$top panel),
the effect of the initial hydrogen abundance, $X_0$, (the right$-$top panel), the effect of core overshooting, $\alpha_{\rm ov}$, (the left$-$bottom panel)
and the effect of rotation, $V_{\rm rot}$, (the right$-$bottom panel).
The following values were adopted for these comparisons: $Z=0.014,~0.020$, $X_0=0.70,~0.75$, $V_{\rm rot}=0, ~0.4 V_{\rm rot}^{\rm crit}$,
where $V_{\rm rot}^{\rm crit}\approx\sqrt\frac{GM}{R}$ is the critical rotation, and  $\alpha_{\rm ov}=0.0, ~0.2$.
The plots are from the age $\log (t/{\rm yr})=5.25$, which corresponds to $t=0.2$ Myr,  because below this value the radius is approximately constant.
The upper age corresponds to an overall contraction phase.

Here, we present results obtained with the OPAL opacities, but as we have checked, the effect of the adopted opacity tables is negligible.

\begin{figure*}
	\includegraphics[width=\textwidth]{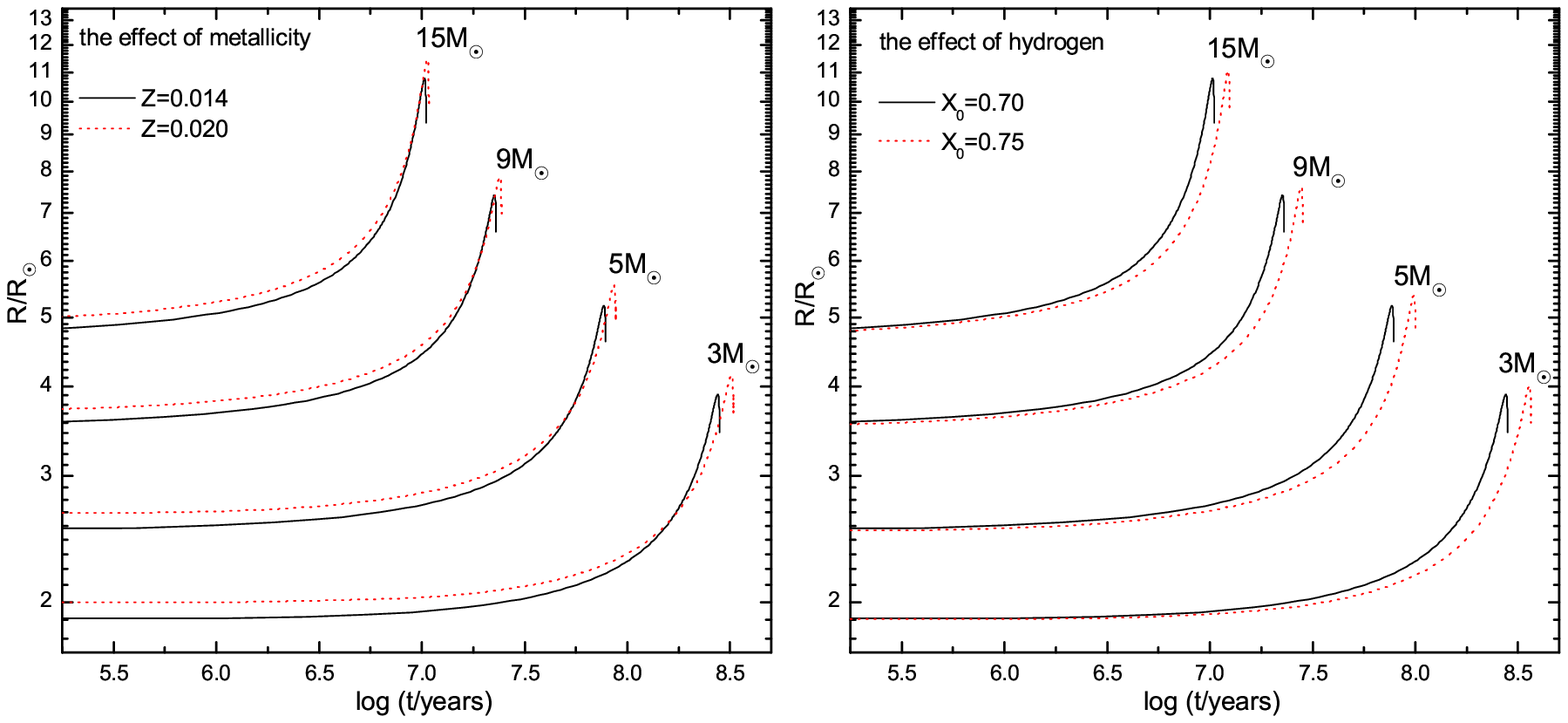}
	\includegraphics[width=\textwidth]{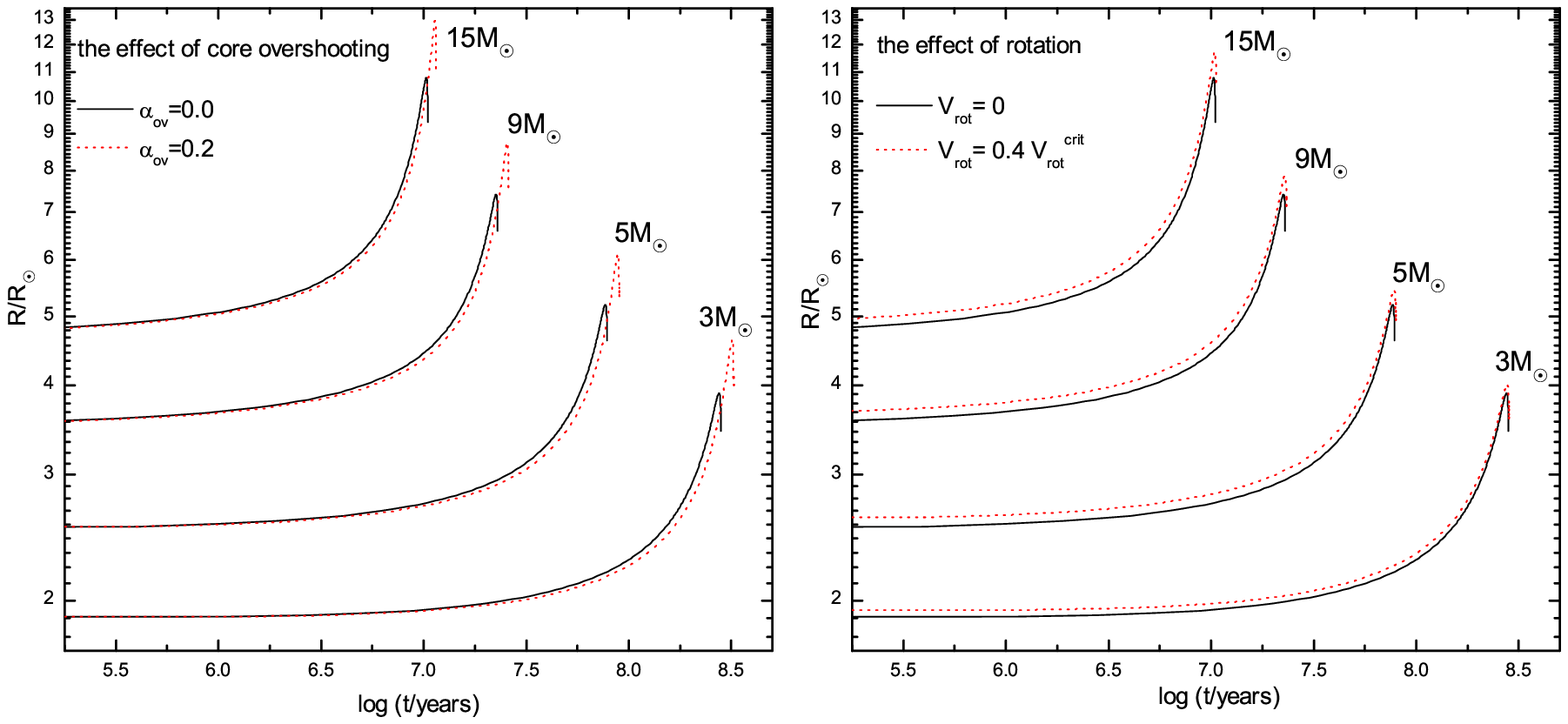}
	\caption{The radius-age diagrams computed for the four values of masses. The top panels show the effect of metallicity
		and initial hydrogen abundance, whereas the bottom panels  - the effect of overshooting from convective core and rotation.
		The tracks are depicted from $\log t =5.25$ ($t\approx 0.2$ Myr) up to the overall contraction phase.}
	\label{fig5}
\end{figure*}

To give a quantitative estimate of theoretical uncertainties, we selected four models with masses 3, 5, 9 and 15 $M_\odot$ (see Table\,2)
and determined the range of age assuming the errors of 3\% both in mass and radius. Then, we investigated the effects of metallicity, $Z$, initial hydrogen abundance, $X_0$,
overshooting from the convective core, $\alpha_{\rm ov}$, and rotation, $V_{\rm rot}$.
We selected the evolutionary stages where differences in the age determination are significant, i.e., all models are roughly in the middle of their main sequence lifespan.
The reference models were computed with the following parameters: $Z=0.014$, $X_0=0.70$, $\alpha_{\rm ov}=0.0$ and $V_{\rm rot}=0$.

\begin{table*}
	\centering
	\caption{{\bf Age estimates with the errors} for models with masses and radii given in the first and second columns, respectively. In the third column, we give the reference age
		determined at metallicity $Z=0.014$, initial hydrogen $X_0=0.70$, without overshooting from the convective core and zero-rotation rate.
		The following columns contain the age estimated when one of these parameters has been changed.}
	\begin{tabular}{c|c|cccccc}
		\hline
		$M$ [$M_{\odot}$]	&  $R$ [$R_{\odot}$]  &	& $\log t$  & $\log t_Z$  & $\log t_X$  &   $\log t_{\rm ov}$   &    $\log t_{\rm rot}$  \\
		$\pm$ 3\%$M$	    &     $\pm$ 3\%$R$	  & &    std    &  $Z=0.020$  & $X_0=0.75$  & $\alpha_{\rm ov}=0.2$ & $V_{\rm rot}^0 = 0.4 V_{\rm crit}$\\
		\hline
		3 		& 2.3 	 & & $8.01\pm0.15$ & $7.93\pm0.19$ & $8.13\pm0.14$ & $8.04\pm0.14$ & $7.37\pm0.46$ \\
		
		5 		& 3.1 	 & & $7.47\pm0.19$ & $7.38\pm0.24$ & $7.58\pm0.18$ & $7.50\pm0.19$ & $6.93\pm0.48$ \\
		
		9 		& 4.4 	 & & $6.97\pm0.12$ & $6.89\pm0.16$ & $7.07\pm0.12$ & $7.00\pm0.12$ & $6.65\pm0.24$ \\
		
		15   	& 6.7 	 & & $6.80\pm0.07$ & $6.76\pm0.07$ & $6.88\pm0.07$ & $6.82\pm0.07$ & $6.66\pm0.09$ \\
		\hline
	\end{tabular}
\end{table*}

In Table\,2, we give the range of age at various sets of parameters for these four selected models. As one can see, increasing metallicity ($Z=0.020$)
will result in a younger age.  However, it should be noted that for models close to Terminal Age Main Sequence (TAMS) the effect is reverse, i.e.,
the higher metallicity gives an older age (see the top$-$left panel of Fig.\,5).
In turn, increasing the abundance of initial hydrogen ($X_0=0.75$) will make always the star older.
Including overshooting from the convective core ($\alpha_{\rm ov}=0.2$) works in the same direction as increasing hydrogen.
On the other hand, increasing the velocity of rotation causes that we will determine the younger age of the star.
The age estimates in Table\,2 can be treated as indicative errors expected from the adopted parameters in evolutionary computations.
At the values of parameters mentioned above and the 3\% errors in mass and radii,
the estimated uncertainties of the age are: $\Delta\log t_Z\approx 0.04-0.09$ for metallicity, $\Delta\log t_X\approx0.11$ for hydrogen abundance,
$\Delta\log t_{\rm ov}\approx 0.02-0.04$ for convective core overshooting and $\Delta\log t_{\rm rot}\approx 0.14-0.64$ for rotation.

However, one has to be aware that there are some correlations between these parameter. 
These correlations can be easily discerned when fitting evolutionary tracks to the position of a star in the HR diagram.
Changing the bulk metallicity $Z$ mimics well the effect of changing a mass, in such a way that increasing the metallicity gives a larger mass
for a fixed position in the HR diagram.
In a similar direction works an increase of the hydrogen abundance $X$ but, in addition, the main sequence is expanded. These effect
will be illustrated in the next section where the results for individual stars will be presented.

The main sequence can be extended also by adding overshooting from the convective core or increasing rotation (cf. Fig\,1).
Higher overshooting shifts evolutionary tracks towards higher effective temperatures and luminosities, thus
a lower mass will be determined for a given position on the HR diagram.
On the contrary, the higher rotation shifts evolutionary tracks towards lower effective temperatures and luminosities.
In this respect, the increase in rotation works in a similar way as increasing the metallicity.

The change of $(X,~Z)$ is limited by the requirement of identical chemical composition for both components
as well as by observational determinations if they exist.
Moreover, the metallcity for most of the stars should not be much different than $Z=0.014$ as determined by \cite{Nieva2012} for galactic B-type stars.
The projected rotational velocities is usually derive
for each component of double-lined eclipsing binaries with a good accuracy (see Table\,1).
Overshooting is described by the free parameter which, of course, can be different for each component.
However, even with these constraints, it may happen that one effect can be compensated with combination of other effects.

\section{The results}

We determine the age assuming that both components are formed at the same time and have the same chemical
composition $(X, Z)$. We based our age determination on the radius-age diagrams plotted for the mass range derived for both components.
Such diagrams are very convenient for the age determination and often used, e.g., recently by \cite{Higl2017}.
The age was determined for each component separately and then the common age was adjusted.
In the next step, the positions of both stars on the HR diagram were confronted with the corresponding evolutionary tracks.

For the vast majority of selected binary stars, the range of the rotational velocity is well constrained by observations
and we use these values. The determination of metallicity [m/H] is available only for seven stars. Usually, this is the value of [Fe/H]
because iron is a good indicator of metallicity and iron lines are prominent and easy to measure.
These determinations are usually not very accurate; with large errors and no errors at all.
Therefore for each binary we  searched the values of {\bf the metal abundance by mass} $Z$ in the range [0.005, 0.030],
which is enough wide for galactic B-type stars with no-peculiar chemical composition. The conversion from [m/H] to $Z$
is given by a simple formula, e.g., \cite{Salaris2005} (Section 8.1, page 239). See also \cite{Eker2018} where there are given
a table and figure allowing to convert easily between $Z$ and [m/H] or [Fe/H].

As overshooting from the convective core strongly affects the lifetime of a star, we also changed the value
of the overshooting parameter, $\alpha_{\rm ov}$. We considered the range $\alpha_{\rm ov}\in[0.0,~0.5]$
and  $\alpha_{\rm ov}$ was adjusted for each value of the metallicity, separately for the component A and B.
We changed the parameters $(Z, \alpha_{\rm ov})$ until the values of mass, radius, effective temperature and luminosity were agreed
with the observational values.  In a few cases, the solution has not been found.

The accuracy of the age determination depends also on the evolutionary stage of the stars. This is because the stellar radius increases slowly
at the beginning of main-sequence evolution, and increases much faster when a star is approaching the end of main sequence.
Therefore, despite of the same accuracy in mass and radius, various accuracy in the age estimation may be reached.

Below, we describe the results for some representative systems for which a consistent solution was obtained
as well as a few problematic cases. The results for all studied systems can be found in Appendix\,A, where
in Table\,3 we give the consistent values of the system age as well as the separate range for each component.
We provide the  solution for $Z=0.014$ and $\alpha=0.0$ if it exists. If not, we give the solution for the values of $(Z,\alpha_{\rm ov})$
that are closest to $(Z,~\alpha_{\rm ov})=(0.014,~0.0)$.
Table \,4 contains the whole range of the parameters $Z$ and $\alpha_{\rm ov}$ for which it was possible to get a common age
from the radius-age dependence and the agreement  on the HR diagram.  The common age from the radius-age relation
and for a specific values of $Z$ and $\alpha_{\rm ov}$ for each component is {\bf listed} in Appendix\,B.

\subsection{Consistent solutions}

For the vast majority of 38 B-type main-sequence binaries it was possible to determine the common age of the two components
as well as to agree their location on the HR diagram. These are the following 33 systems:
AH Cep, V578 Mon, V453 Cyg, V380 Cyg, CW Cep, NY Cep, V346 Cen, DW Car, V379 Cep, QX Car, V399 Vul, V497 Cep,
V539 Ara, CV Vel, LT CMa, AG Per, U Oph, DI Her, EP Cru, V760 Sco, MU Cas, GG Lup, V615 Per,  BD+03 3821, V1665 Aql,
$\zeta$ Phe, YY Sqr, V398 Lac, $\chi^2$ Hya, V906 Sco, $\eta$ Mus, V799 Cas, PV Cas.
In the case of V380 the consistent solution was possible only if the primary was in the post-main sequence phase after an overall contraction.

For some binaries we had to change the metallicity or add overshooting from the convective core
to agree the star's positions on the HR diagram. In some cases, both of these parameters had to be changed.

For example, for MU Cas the observed metallicity derived  by \cite{Lacy2004} is $Z=0.025$.
With this value of $Z$ it was possible to determine the common age from the radius-age relation but it gave a disagreement in the position
of the stars on the HR diagram. It appeared that the solution with $Z=0.014$ is fully consistent.
The other cases which required changing metallicity were: V453 Cyg, DW Car, V379 Cep, V399 Vul, LT CMa, DI Her, V615 Per, BD+03\,3821,
V1665 Aql,  $\zeta$ Phe, V398 Lac, $\chi^2$ Hya, V906 Sco, $\eta$ Mus and PV Cas.

Another example is the binary V615 Per. This is a member of the young open cluster h Persei (NGC 869) for which the estimated age from photometric observations
is $13 -14$ Myr \citep[e.g.,][]{Capilla2002, Currie2010}. However, this value was obtained assuming a solar metallicity.
For the metallicity higher than $Z=0.009$ and  $\alpha_{\rm ov}=0.0$ it is possible to determine only the upper limit of the age
because the curves $R(\log t)$ do not intersect the line $R=R_{\rm min}$
On the other hand, the value of at least $Z=0.009$ was needed to agree the positions of both components on the HR diagram.
We got the age range $0.8 - 32.0$ My. In order to narrow down this range, it would be necessary to reduce the errors of radius determinations,
which amount to about 6\%.

The metallicity of $Z\approx 0.01$ for V615 Per was suggested by  \cite{Southworth2004A} from reproducing masses and radii with the evolutionary computations. But the authors reduced the abundance of hydrogen to $X=0.63$ to keep the age of 13 Myr.
The subsolar metallicity was derived also from fitting disentangled spectra of V615 Per by \cite{Tamajo2011}.

Below we give details of the subsequent steps of the age determination for the three systems. We selected binaries with masses around
10\,$M_{\sun}$,  6\,$M_{\sun}$ and 3\,$M_{\sun}$.

\vspace{0.25cm}
{\bf V578 Monocerotis}
\vspace{0.15cm}

\begin{figure}
	\includegraphics[width=\columnwidth, clip]{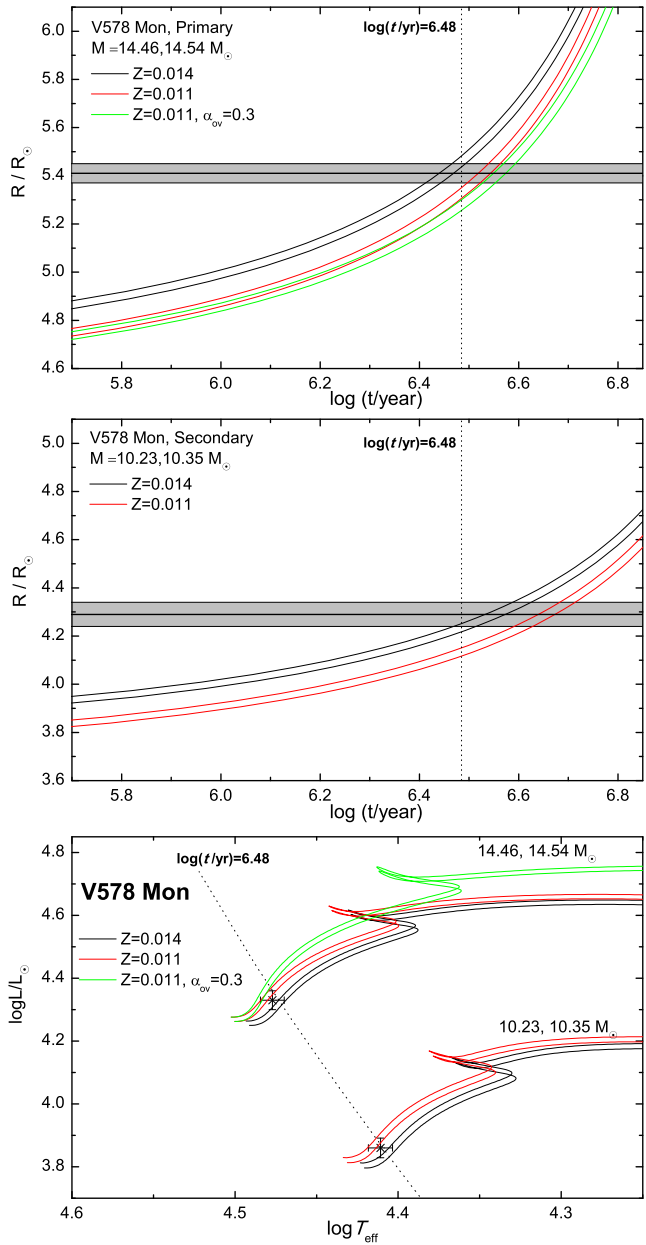}
	\caption{The top and middle panels show evolutionary changes of the radius for the minimum and maximum mass of the primary and secondary component
        of V578 Mon, respectively. There is shown the effect of metallicity and the effect of core overshooting for the primary.
		The bottom panel shows the HR diagram with the location of the two components. 
}
	\label{fig6}
\end{figure}

V578 Mon is one of the few double-lined eclipsing binaries with masses of both components above 10\,$M_\odot$
and with masses and radii determined with an accuracy below 3\%. Therefore, the object has been already studied by several groups.
The recent determination of basic stellar parameters was done by \cite{Garcia2014} and we used their values of masses and radii in this work, i.e.,
$M_A=14.54(8)~M_{\sun}$, $R_A=5.41(4)~R_{\sun}$ for the primary and $M_B=10.29(6)~M_{\sun}$, $R_B=4.29(5)~R_{\sun}$ for the secondary.
Comparing to the values of \cite{Torres2010}, the errors given by \cite{Garcia2014} are about 30\% lower in mass and 50\% lower in radius.
The star is a member of the OB cluster NGC\,2244 which drives the star-forming region called the Rosette nebula \citep[e.g.,][]{Roman-Zuniga2008}.
The metallicity of V578 Mon from disentangled component spectra is [m/H]=-0.30(13) \citep{Pavlovski2005} and for the whole cluster NGC\,2244
\cite{Tadross2003} estimated [Fe/H]$\approx$-0.46. The age of the binary derived from disentangled component spectra by \cite{Hensberge2000}
is 2.3(2) Myr ($\log (t/{\rm yr})=6.36(4)$.
The estimated age of NGC 2244 from 2MASS photometry by \cite{Bonatto2009} is in the range $1 - 6$ Myr ($\log (t/{\rm yr})=6.00 - 6.78$).

On the other hand, the determination of age of V578 Mon involving more advanced evolutionary modelling did not give consistent result.
\cite{Garcia2014} did not find a common age of both components and invoked a large overshooting from the convective core
for the primary to lower the discrepancy. Recently, \cite{Higl2017} obtained similar results for this binary.
They also showed that the assumption of solar metallicity, rather than [Fe/H]=-0.3 as determined from observations, reduces the age gap between components.
This age discrepancy for the V578 Mon system was obtained also by \cite{Schneider2014} in the framework of the BONNSAI project.

Contrary to the previous works, our models allowed to find the consistent age the V578 Mon system,
i.e., the common age from the radius-age relation and the agreement in the star's positions on the HR diagrams.
At the metallicity $Z=0.011$, which corresponds approximately to [m/H]=-0.2, and the overshooting of at least $\alpha_{\rm ov}=0.3$
for the more massive component, we determined the age of about 3.89 Myr ($\log (t/{\rm yr})=6.59$). At the metallicity $Z=0.014$
there is no need to include overshooting and the age of the system is 3.02(7) Myr ($\log (t/{\rm yr})=6.47-6.49$).
The consistent age exists for $Z\in[0.011,0.016]$ and various combinations of the overshooting parameters for both components
which are in the ranges: $\alpha_{ov}(A)\in[0.0,~0.5]$ and $\alpha_{ov}(B)\in[0.0,~0.2]$.

The method of determining the age is illustrated in Fig.\,6, where we depicted the evolution of the radius for the minimum and maximum mass allowed
by observations. The top and middle panels correspond to the primary and secondary component, respectively.
The horizontal grey belts mark the observed ranges of the radii.
As one can see both solutions, with $Z=0.011$ and $Z=0.014$, are consistent with the positions of the components on the HR diagrams shown
in the bottom panel of Fig.\,6.  With dotted line we depicted also the isochrone $\log (t/{\rm yr})=6.48$ for $Z=0.014$.

The case of V578 Mon demonstrates very well the correlation between
the overshooting parameter and metallicity. Therefore in Table\,3, we give as an example both solutions, for $Z=0.011$ and $Z=0.014$.

The first solution is more consistent with the metallicity determined from spectroscopy by \cite{Pavlovski2005}. The larger parameter of overshooting
for the more massive primary could be also explained by the need of some additional mixing which results not necessarily from overshooting itself
but, for example, from diffusion or/and rotation.
The second solution, without overshooting, requires higher metallicity than the observed value, i.e., about $Z=0.014$,
which is, in turn, a typical value for galactic B-type stars \citep{Nieva2012}.

Both our estimates of age are more or less consistent with the previous determinations for the cluster NGC 2244.

\vspace{0.25cm}
{\bf CV Velorum}
\vspace{0.15cm}

The system of CV Vel consists of two B-type stars with almost the same masses determined with a very high accuracy: $M_A=6.086(44)~M_{\sun}$ and $M_B=5.982(35)~M_{\sun}$,
what gives the errors below 0.8\% in mass. The values of radii are determined with a similar accuracy: $R_A=4.089(36)~R_{\sun}$ and $R_B=3.950(36)~R_{\sun}$.
Several estimates of age can be found in the literature for this binary. The oldest determination by \cite{Clausen1977} gives about 30 Myr ($\log (t/{\rm yr})=7.48$).
From the membership of CV Vel in IC2391, \cite{Gimenez1996} estimated the age between 33 and 82 Myr ($\log (t/{\rm yr})=7.72\pm 0.20$), whereas,
based on evolutionary models, \cite{Yakut2007} derived 40 Myr ($\log (t/{\rm yr})=7.60$).
The most recent determination by \cite{Schneider2014} is about 35 Myr.

In the top and middle panels of Fig,\,7, we show the evolution of radius for the two components considering two values of metallicity, $Z=0.014$ and $0.010$.
The lower value of $Z$ was suggested by \cite{Yakut2007}. For both metallicities it was possible to determine the common age of the system from such diagrams, but, as one can see in the HR diagram (the bottom panel of Fig.\,7), only the higher metallicity solution is consistent with the observational error boxes.

We derived $t=31.6 - 33.9$ Myr ($\log (t/{\rm yr})=7.50 -7.53$) for $Z=0.014$. The isochrone $\log (t/{\rm yr})=7.515$ for $Z=0.014$ is marked in Fig.\,7.
The consistent solution was found for the metallicity $Z\in[0.011,~ 0.018]$ and the overshooting parameters $\alpha_{ov}\in[0.0,~0.2]$ for both components.

\begin{figure}
	\includegraphics[width=\columnwidth, clip]{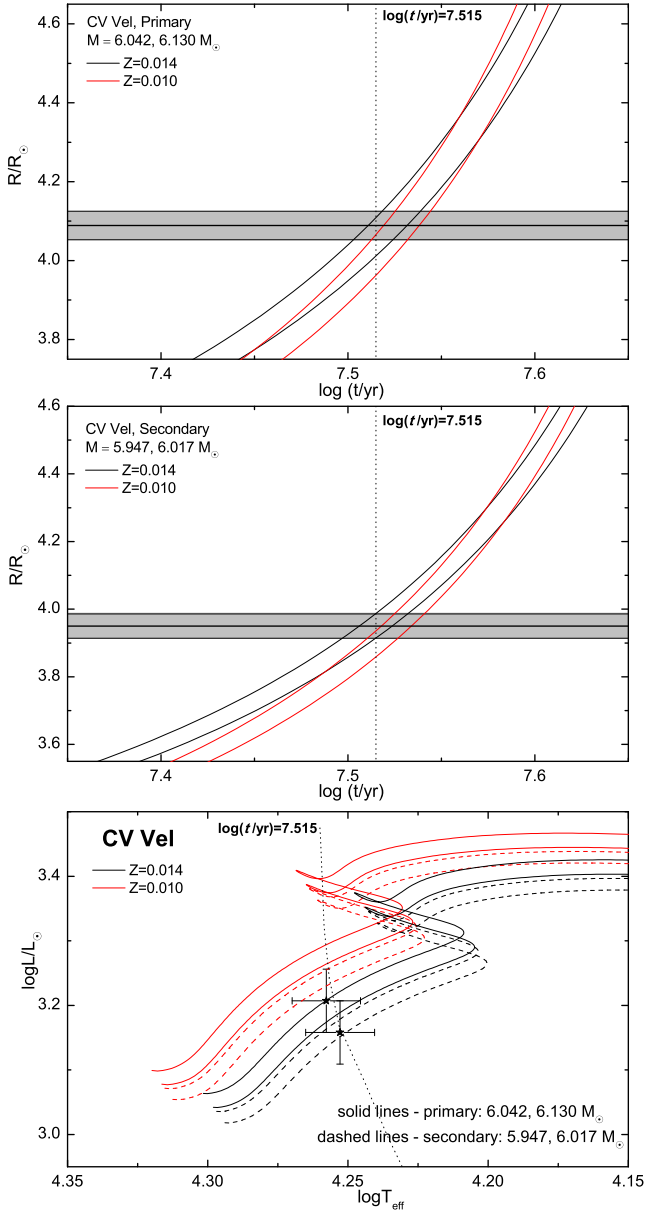}
	\caption{The similar plots as in Fig.\,6 but for the CV Vel system.}
	\label{fig7}
\end{figure}

CV Vel was also a subject of the BANANA project by \cite{Albrecht2014}. They found the misalignment of the rotation axis of the primary star
and the changes of the projected rotation velocity of both stars. The latter finding can be interpreted as precession of the rotation axes \citep{Albrecht2014}.
It is worth to mention also that \cite{Yakut2007} reported line profile variability which could be related with pulsations.
This is very plausible because both components are located in the instability strip of Slowly Pulsating B-type stars.
Detection of pulsational frequencies would allow for more in-depth studies of the system by mean of asteroseismic modelling.

\vspace{0.25cm}
{\bf V906 Scorpii}
\vspace{0.15cm}

V906 Sco is a triple system in the open cluster NGC 6475 (M7) \citep{Alencar1997}. The eclipsing binary is composed of late B-type stars with masses:
$M_A=3.378(71)~M_{\sun}$ and $M_B=3.253(69)~M_{\sun}$, and radii: $R_A=4.521(35)~R_{\sun}$ and $R_B=3.515(39)~R_{\sun}$. The third component has a similar spectral type (B9V)
and is on the wide orbit with the period of the order of hundred years \citep{Alencar1997}. Its contribution to the total light is about 5\%.

The metallicity of NGC\,6475 estimated by \cite{Sestito2003}, is [Fe/H] = $+0.14(6)$ whereas \cite{Villanova2009} derived
[Fe/H] = $+0.03(2)$ and suggested oversolar helium abundance $Y=0.33(2)$. The age of the cluster is about 220 Myr according to \cite{Meynet1993}
and 200$\pm$50 Myr according to \cite{Villanova2009}.

Fig.\,8 shows the run of $R(\log t)$ for the allowed mass range of the two components of the V906 Sco binary (the top and middle panels)
and their location on the HR diagram (the bottom panel). Following the estimates in the literature, we considered three compositions:
($X_0=0.70,~Z=0.014$), ($X_0=0.70,~Z=0.017$) and ($X_0=0.65,~Z=0.015$).
The primary component is quite evolved stars; it is already close to TAMS and overshooting of at least $\alpha_{\rm ov}=0.1$
is required for the star to be on main sequence for each values of ($X_0,~Z$). In Fig.\,8 we depicted the evolution of the radius for $\alpha_{\rm ov}=0.2$.
No overshooting is needed for the secondary star.
The common age from the radius-age diagrams
was determinable in each case and we obtained 208.5(1.5)  Myr  ($\log (t/{\rm yr})=8.316 -8.322$)  for ($X_0=0.70,~Z=0.014$),
$\sim$219 Myr ($\log (t/{\rm yr})=8.340 - 8.341$)  for ($X_0=0.70,~Z=0.017$)
and 164(1) Myr ($\log (t/{\rm yr})=8.213 - 8.217$)  for ($X_0=0.65,~Z=0.015$).
Here, we got a younger age for lower metallicity because the primary star is close to TAMS.

The evolutionary tracks used to determine the age are depicted on the HR diagram in the bottom panel of Fig.\,8 together with error boxes for both components. The isochrone $\log (t/{\rm yr})=8.319$ for $X_0=0.70,~Z=0.014$ and $\alpha_{\rm ov}=0.2$ is depicted as the dotted line.
As one can see, only solutions with the initial hydrogen abundance $X_0=0.70$ are consistent and the corresponding evolutionary tracks
are within the error boxes of both stars.
Moreover, in these cases our estimates of the age for  $\alpha_{\rm ov}(A)=0.2$ ($\sim209$ Myr for $Z=0.014$ and $\sim219$ for $Z=0.017$)
are in agreement with the age of NGC6475.
In turn, the solution with the hydrogen abundance of  $X_0=0.65$, as suggested by \cite{Villanova2009}, gives inconsistent values
of luminosity and the system is too young.

The similar result was obtained by \cite{Higl2017} who derived the age of 236(13) Myr, which is slightly higher that our values.

\begin{figure}
	\includegraphics[width=\columnwidth, clip]{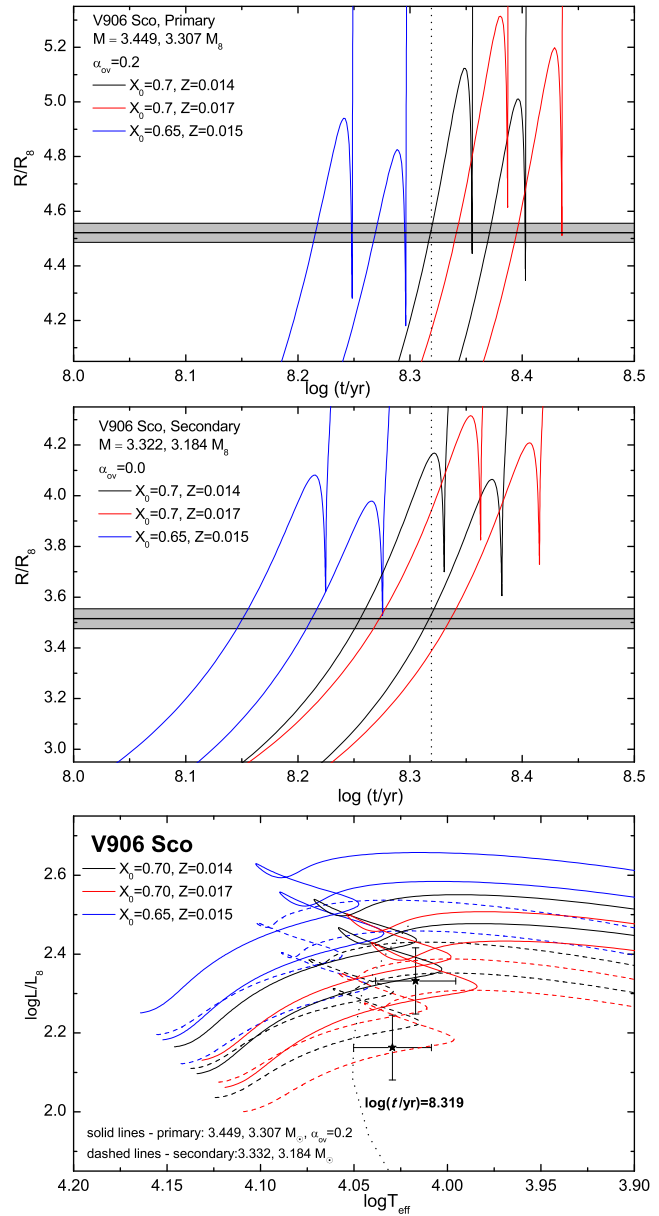}
	\caption{The similar plots as in Fig.\,6 but for the V906 Sco binary.}
	\label{fig8}
\end{figure}

\subsection{Problematic cases}

There were a few systems for which it was impossible to get consistent solutions, i.e., to agree the age of both components from the radius-age relation
and their positions on the HR diagram. These are the binaries:   HI Mon, V380 Cyg, V1331 Aql, V1388 Ori, IM Mon and V413 Ser.

We did not determine the age for HI Mon because of a disagreement on the HR diagram.
To determine the common age for both components from the radius-age diagram a large overshooting $\alpha_{\rm ov}>0.5$ for the primary was indispensable,
but then the primary was underluminous for its mass.
HI Mon is a massive binary ($M_A=14.2(3)~M_{\sun}$ and $M_B=12.2(2)~M_{\sun}$) with both components rotating
quite fast ($V_{\rm rot}\approx 150$ km/s). Thus, maybe other effects, like differential rotation or rotation-induced mixing
have to be taken into account.

For one system, V1331 Aql, we were unable to determine even a common age of both components from the radius-age relation.
There is also a strong disagreement in the position on the HR diagram for the secondary,
which is oversized and overluminous for its mass.
Changing any parameter in a reasonable range does not lead to a consistent solution.
As show by \cite{Lorenz2005}, the secondary of V1331 Aql reaches about 96\% of the Roche radius.
Thus, V1331 Aql is a detached configuration but the secondary is about to fill its Roche lobe.

It was not possible to get the consistent age for the binary V413 Ser because of a disagreement on the HR diagram
for all searched values of $Z$ and $\alpha_{\rm ov}$. V413 Ser is located in the Serpens star-forming region \citep{Chavarria1988}.
The Serpens dark cloud contains an extremely young star cluster and most probably both components of V413 Ser are pre-main sequence objects \citep{Cakirli2008}, and not the main-sequence ones as we assumed in our age determination.

Results for IM Mon, Ori V1388 and V380 Cyg will be presented below.

\vspace{0.25cm}
{\bf IM Monocerotis}
\vspace{0.15cm}

Till now, IM Mon was studied in details only in two papers, i.e., by \cite{Bakis2010} and \cite{Bakis2011}.
In \cite{Bakis2011}, the analysis of all available photometric and spectroscopic observations allowed to determine
the absolute values of masses: $M_A=5.35(24)~M_{\sun}$ and $M_B=3.32(16)~M_{\sun}$,
and radii: $R_A=3.15(4)~R_{\sun}$ and $R_B=2.36(3)~R_{\sun}$.
Moreover, it has been shown that IM Mon is a member of Ori OB1a association
and the age of the system is about 11.5 Myr. The age was estimated from a comparison of the location of stars
on the diagram $\log T_{\rm eff} - \log g$ with the isochrones of \cite{Girardi2000}.
A comparison of the observed high-resolution spectrum with Kurucz atmosphere models  gave atmospheric parameters and
metallicity [Fe/H]$=+0.2$.
The age of IM Mon determined by \cite{Bakis2011}  is in agreement with earlier determinations for Ori OB1a association
by \cite{Blaauw1991} and \cite{Brown1999}.

Using the masses and radii, we tried to re-determine the age of the system, as illustrated in Fig.\,9. We found that metallicity of at least $Z=0.025$
is needed to find a common age of the components which is in line with the values of [Fe/H] obtained by \cite{Bakis2011}.
Then, the age of the system is about 9.7(2.9) Myr, which is in agreement with the previous determinations.
However, when we put the stars on the HR diagram (the bottom panel of Fig.\,9),
it turned out that the secondary component is overluminous for its mass.
The evolutionary tracks for the metallicity typical for galactic B-type stars, $Z=0.014$, would be consistent, although marginal,
but in this case there is no common age for both components even if the large overshooting ($\alpha_{\rm ov}=0.5$) is assumed for the primary star.
One reason of that could be too hight effective temperature. The lower value of $T_{\rm eff}$ would move the star
to the right and down in the HR diagram.
The second explanation could be a more complicated configuration of this binary, i.e, maybe the system is not fully detached
or there is a third light.
\begin{figure}
	\includegraphics[width=\columnwidth, clip]{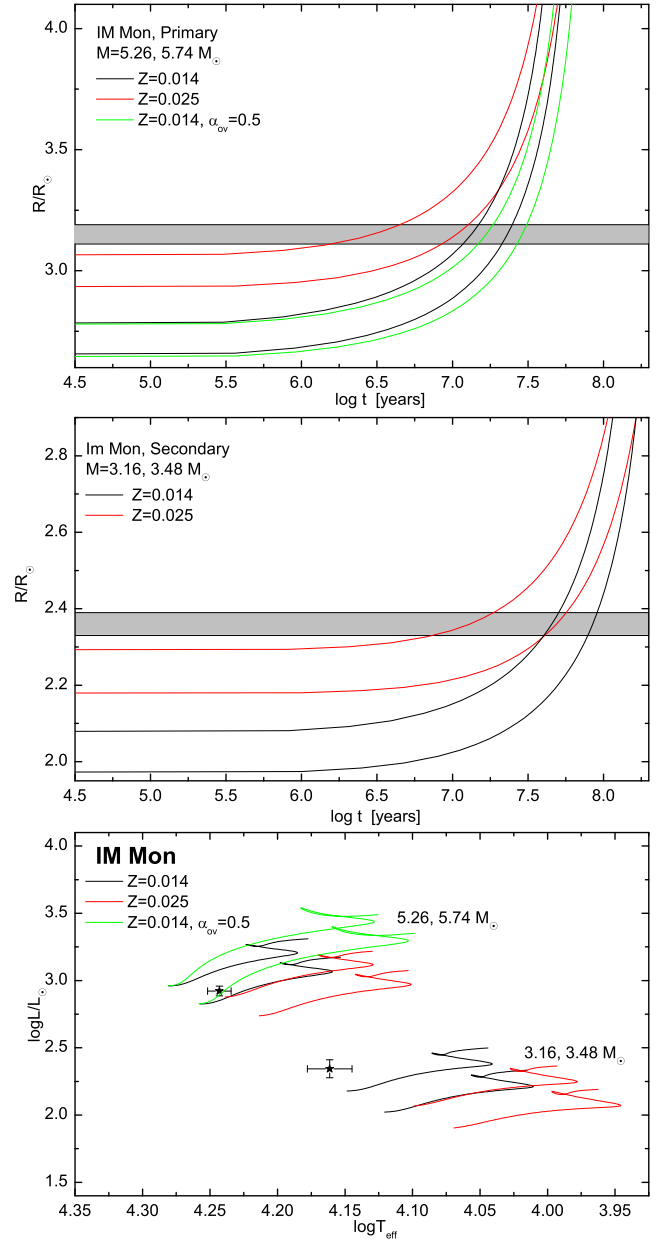}
	\caption{The similar plots as in Fig.\,6 but for components of IM Mon.}
	\label{fig9}
\end{figure}
\begin{figure}
	\includegraphics[width=\columnwidth, clip]{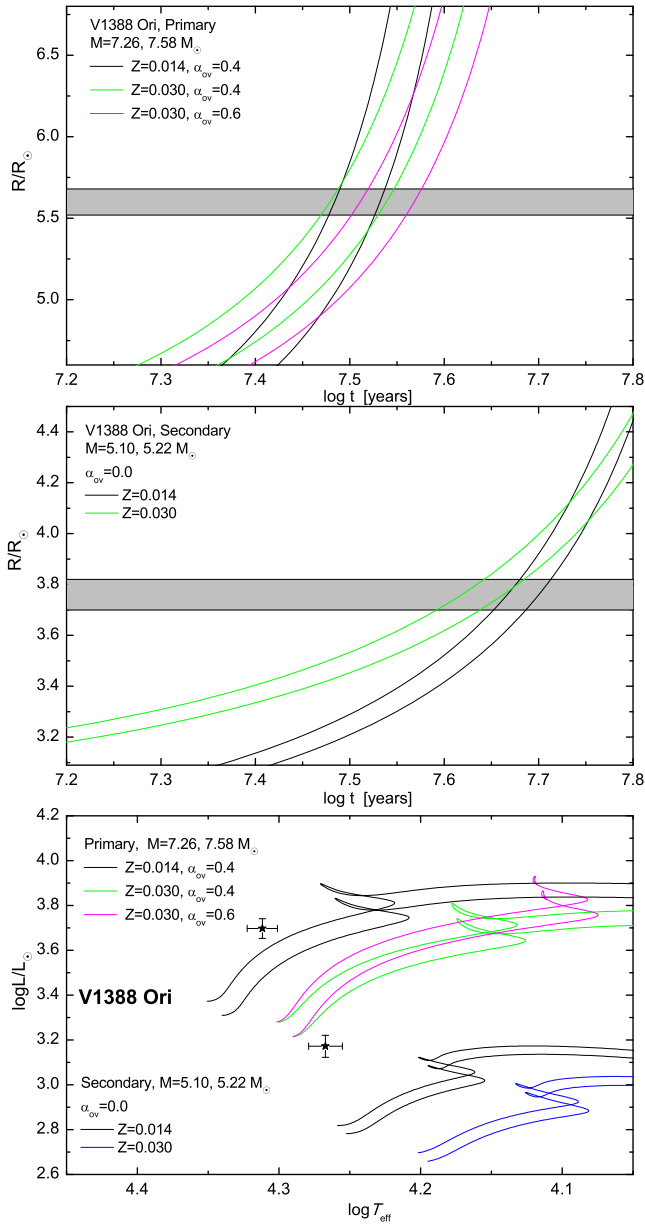}
	\caption{The similar plots as in Fig.\,6 but for V1388 Ori. }
	\label{fig10}
\end{figure}

\vspace{0.25cm}
{\bf V1388 Orionis}
\vspace{0.15cm}

This binary is located close to Galactic open cluster NGC 2169 but because of significant age difference its cluster membership is unlikely.
The absolute values of masses and radii of the components of V1388 Ori were determined by \cite{Williams2009} and they are
$M_A=7.42(16)~M_{\sun}$, $M_B=5.16(6)~M_{\sun}$, and $R_A=5.60(8)~R_{\sun}$, $R_B=3.76(6)~R_{\sun}$.
From the position on the HR diagram, \cite{Williams2009} set
the age of the system at about 25 Myr, whereas the age of NGC 2169 is about $9\pm 2$  Myr as estimated by \cite{Jeffries2007}.
Fitting the spectral energy distribution to the photometric measurements in the passband $U BVJHK_s$ transformed to fluxes
allowed to estimate the distance $d=832\pm 89$ pc.
Up to now, no common age was determined for the components of the V1388 Ori system \citep[e.g.,][]{Schneider2014}.

We have attempted to agree the age of both components by changing the metallicity and the parameter of core overshooting.
Adopting the metallicity $Z=0.014$ did not lead to a consistent solution; always the primary was much younger than the secondary.
Even the high values of the overshooting parameter $\alpha_{\rm ov}>0.4$ for the primary did not shift the age to the enough higher values.
The corresponding curves $R(\log t)$  are plotted in the top and middle panels of Fig.\,10.

It was possible to make the primary older and the secondary younger by increasing the metallicity up to $Z=0.03$. This is because the primary
is close to TAMS and the effect of metallicity is reverse, i.e., the higher the metallicity the older the age of a star.

Comparing the positions of the V1388 Ori stars on the HR diagram with evolutionary tracks confirmed the previous results in the literature
that in the case of this binary both components are much overluminous for their masses \citep{Williams2009}. The discrepancy is stronger
for higher metallicity because for a given mass the evolutionary tracks are shifted to lower effective temperatures and lower luminosities.

\vspace{0.25cm}
{\bf V380 Cygni}
\vspace{0.15cm}

The binary V380\,Cyg consists of two early B-type stars with a significant mass difference ($M_B/M_A\approx0.6$).
There are two most recent determinations of masses and radii of these system components.
\cite{Pavlovski2009} derived the following values of masses: $M_A=13.13(24)~M_{\sun}$ , $M_B=7.78(10)~M_{\sun}$,
and radii: $R_A=16.22(26)~R_{\sun}$ and $R_B=4.060(84)~R_{\sun}$.  Lower masses and radii were derived by \cite{Tkachenko2014}
who used high-precision photometry obtained by the Kepler space mission and high-resolution ground-based spectroscopy.
Below, we first provide details of the age determination adopting masses and radii of \cite{Pavlovski2009}.

Assuming the metallicity of $Z=0.014$  and no-overshooting from the convective core, the primary is beyond the main sequence,
as can be seen from the top and bottom panel of Fig.\,11.
Besides, the primary is much overluminous comparing with the corresponding evolutionary tracks.
To catch the massive component on the main sequence or just after it,
the overshooting from the convective core of at least $\alpha_{\rm ov}=0.5$ was indispensable. Such value was
suggested for the first time by \cite{Guinan2000}.
We tried to determined the age of V380 Cyg in the three phases of evolution: 1) main-sequence, 2) post-main sequence in an overall contraction,
and 3) post-main sequence after an overall contraction. In each case, it was possible to find the common age from the radius-age relation.
However, only in the third case (after the overall contraction) the agreement in the positions on the HR diagram has been reached.
In Table\,3, we provide this solution.

Adopting the masses and radii determined by \cite{Tkachenko2014}, it was not possible to derive the consistent age for any phase of evolution
because of a disagreement on the HR diagram for the primary.

Based on current observations, we cannot resolve what is the evolutionary stage of the V380 Cyg system and how our solution after the overall contraction is reliable.
Of course, observing a star in the main-sequence phase is much more likely due to time scales, but the probability of catching even a quite massive star in the post-main sequence
phase is not completely zero.
All scenarios demand a large overshooting parameter. As in the case of V578 Mon, the question is whether such effective overshooting
from the convective core has any physical origin. In particular: is the mixing at the core boundary really so large, or maybe a large overshooting parameter
compensates for the effects of other mixing processes?

More observations and analysis {\bf are} needed to answer the question about the evolutionary stage of the V380 Cyg primary.
Here, the studies of pulsations of the primary star would help but contrary to \cite{Tkachenko2014}, we did not find any oscillation frequencies in the Kepler light curve \citep{Miszuda2018}.

Similarly, computations with the more advanced evolutionary code MESA  
did not resolve the evolutionary stage of the massive component of V380 Cyg \citep{Tkachenko2014, Miszuda2018}.

\begin{figure}
	\includegraphics[width=\columnwidth, clip]{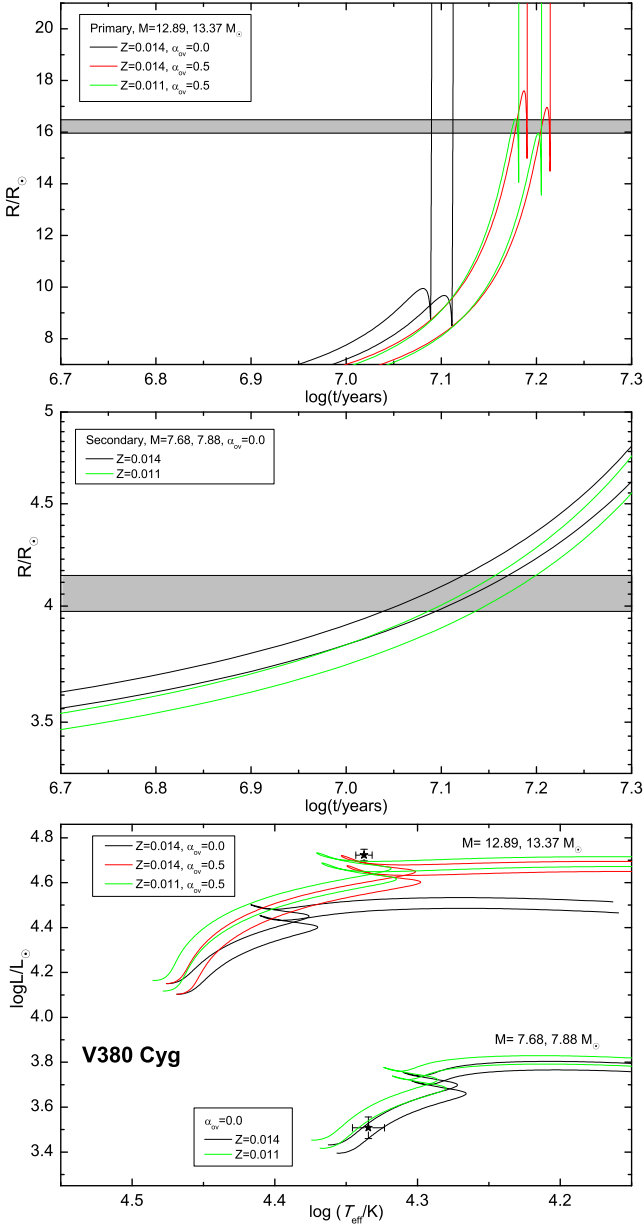}
	\caption{The similar plots as in Fig.\,6 but for V380 Cyg.}
	\label{fig11}
\end{figure}

\subsection{Discussion}

For most binaries, an equal age solution was found from the mass-radius-age diagram that is also compatible with the effective temperature
and luminosity of individual stellar components. In some case, the additional adjustment of metallicty or the overshooting parameter was necessary.
There are also the systems where an equal age solution is unsatisfactory because
it was not possible to agree positions of individual stellar components in the HR diagram.
For example, in the binary IM Mon, the secondary was overluminous and
in the binary V1388 Ori both components were overluminous. Another, completely different, problematic case was the V380 Cyg system
in which the evolutionary stage of the primary is uncertain. The primary star can be still on the main-sequence or has just entered a post-main sequence evolution
but the consistent age exists only if the primary is the post-main sequence star after the overall contraction.

Besides, as mentioned already, the accuracy of the derived age depends on the evolutionary stage and even if masses and radii are determined
with very small errors (below 3\%) the age cannot be determined accurately; an example is the binary DI Her.

The ages of the primary and secondary stars of binaries are plotted against each other in Fig.\,12.
As one can see, only V1331 Aql is an outlier taking into account the errors.
%
\begin{figure}
	\includegraphics[width=\columnwidth, clip]{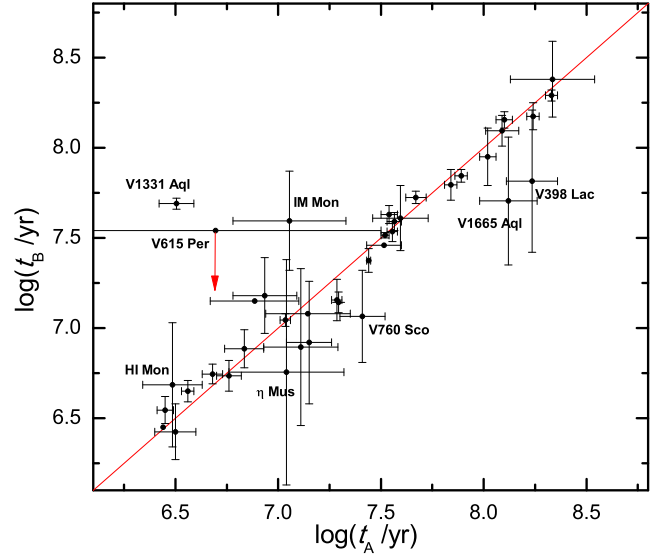}
	\caption{The age of the primary and secondary component from the radius-age relation for all 38 binaries from our sample.}
	\label{fig12}
\end{figure}

A comparison of the observed values of the effective temperatures, $T_{\rm eff}$ and luminosities, $\log L/L_\odot$, with those derived
from evolutionary computations ("evol") is presented in Fig.\,13. Only 33 binaries with the consistent age was included.
The left-hand panel correspond to the effective temperature, $T_{\rm eff}$, and the right-hand panel to the luminosities, $\log L/L_\odot$.
As one can see the agreement between the observed and evolutionary values is very good. This indicates that the predicted theoretical stellar tracks
and isochrones would also agree with the star's position on the HR diagram.

\begin{figure*}
	\includegraphics[width=\columnwidth, height=7.5cm, clip]{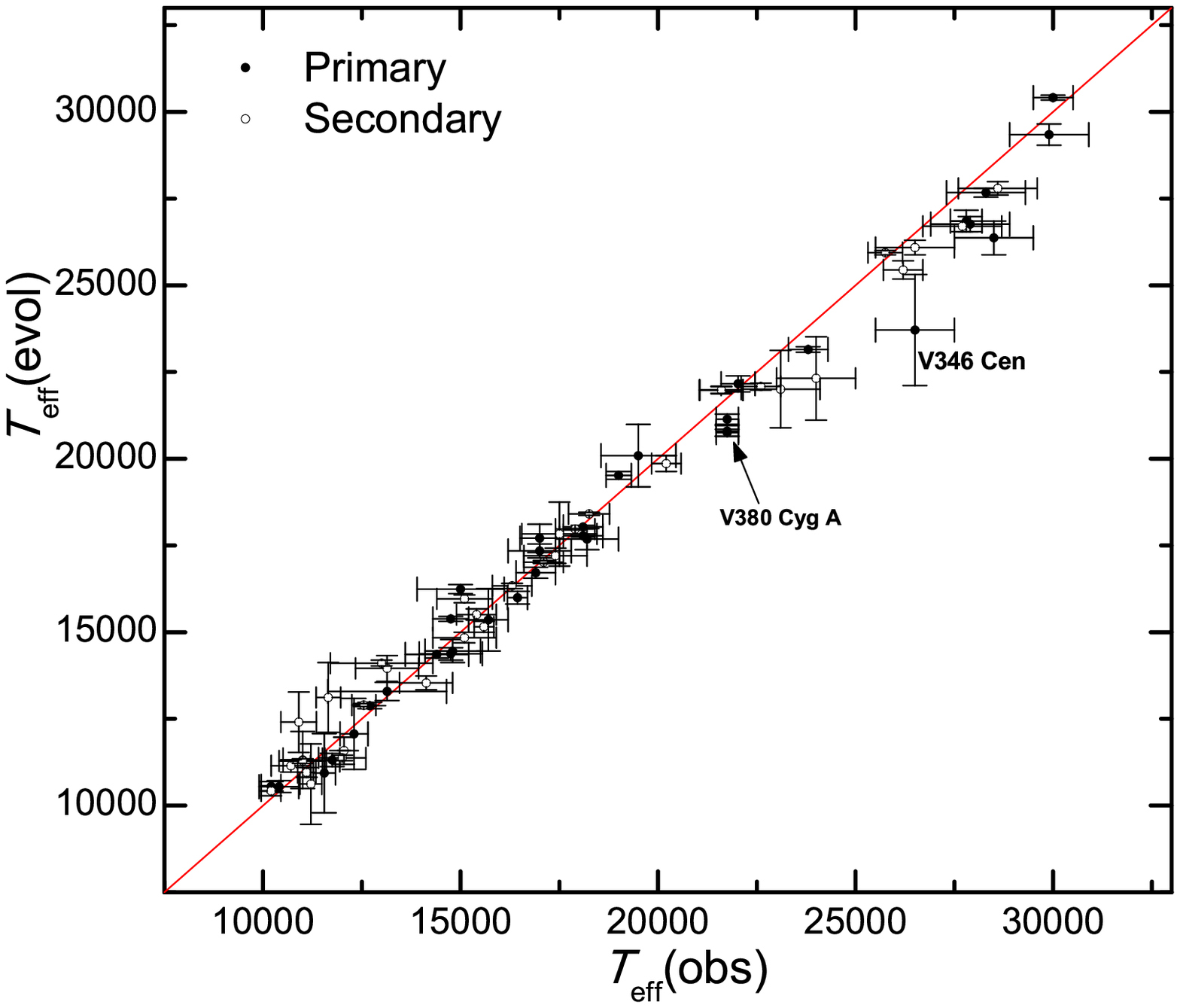}
	\includegraphics[width=\columnwidth, height=7.5cm, clip]{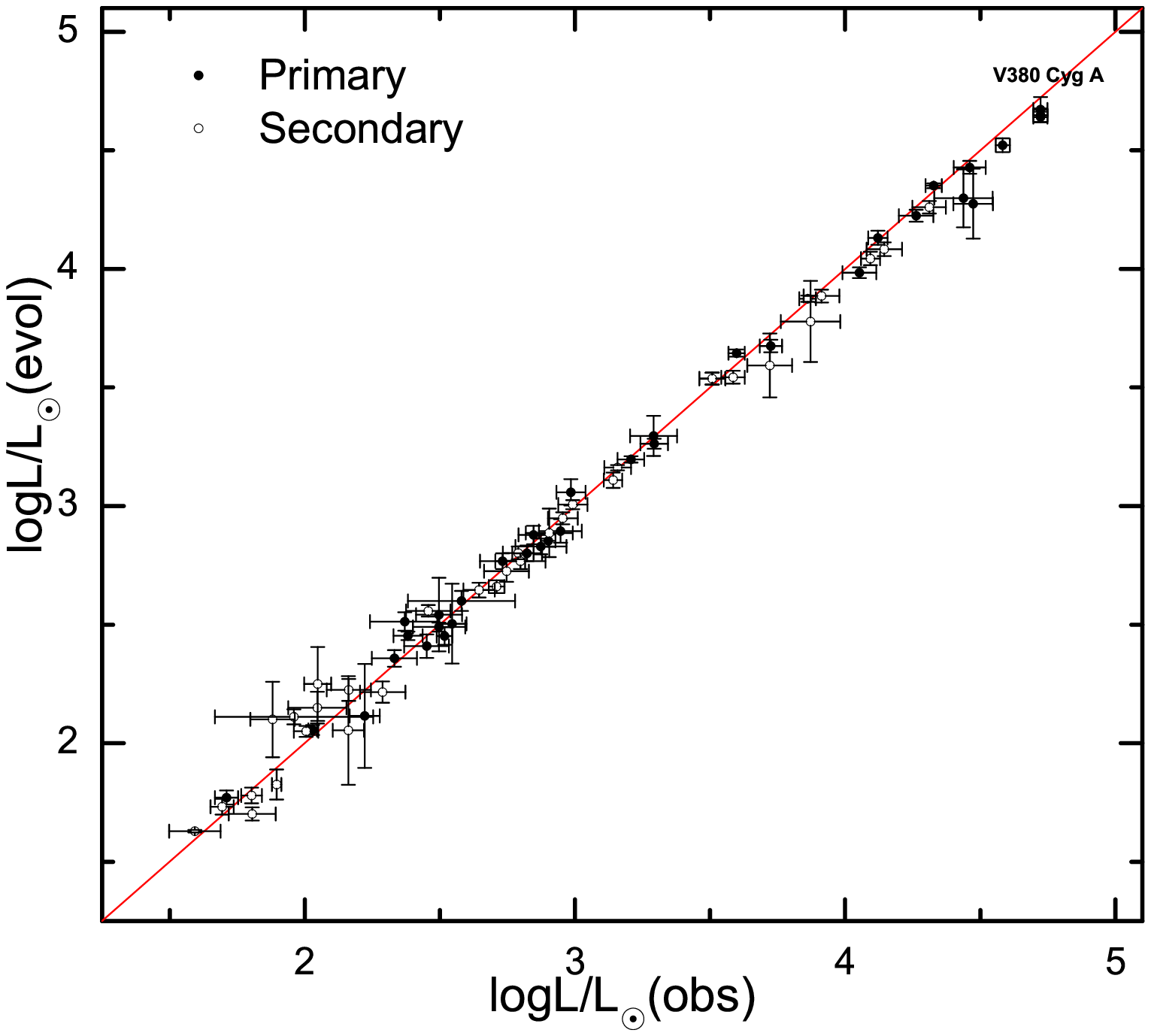}
	\caption{A comparison of the values of the effective temperature (the left panel) and luminosity (the right panel)
		determined from observations with the corresponding values derived from evolutionary models reproducing the consistent age
		of the binary systems.}
	\label{fig13}
\end{figure*}

Fig.\,14 shows a comparison of our age determination with the values gathered from the literature.
These values are also listed in the Appendix\,C in Table\,5 with the references to their sources.

There is one clear outlier, AG Per, which is generally accepted as a member of the Per OB2 association \citep{Blaauw1952}.
The age determined by \cite{Gimenez1994} is about 50(10) Myr which is much above our value, 15.6(2.5) Myr.
Our estimation is, however, compatible (within the error) with the age of Per OB2
which, according to \cite{Gimenez1994}, amounts to about 12.5(2.5) Myr.

Some disagreement is also for the four more stars: GG Lup, NY Cep, $\zeta$ Phe and PV Cas.
It can results from different approach to the age determination.
In these cases, the literature values are rather rough estimates, without errors given.

For other stars our values of the age are consistent with those found in the literature,
even if the latter are just rough estimates as in cases of V359 Ara, CV Vel, U Oph, V1665 Aql, YY Ser, $\eta$ Mus and V799 Cas.

\begin{figure}
	\includegraphics[width=\columnwidth, clip]{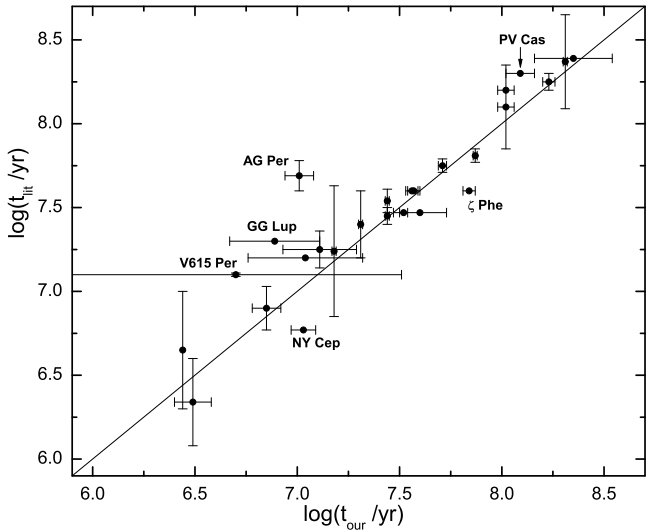}
	\caption{Comparison of the age of B-type binaries determined in this paper with the literature values.}
	\label{fig14}
\end{figure}

\section{Summary and future studies}

The goal of this paper was to determine the age for a homogeneous sample of stars. To this aim we chose main-sequence B-type stars in double-lined eclipsing binaries.
There are 38 suitable systems and for most of them the masses and radii are determined with the accuracy below 3\%.
These stars are relatively simple objects because of the lack of subsurface convection and no significant mass loss.
Thus, in evolutionary modelling we get rid of two additional free parameters describing these phenomena.
On the other hand, the B-type stars are very important for evolution of chemical elements and structure of galaxies.
Moreover, the properties of main-sequence objects are fundamental for further evolution and ultimate fate of stars.

In all cases but one it was possible to determine the common age for both components from the radius-age diagrams. 
The exception is the system V1331 Aql in which the secondary is about to fill its Roche lobe.
There were also five systems, HI Mon, IM Mon, V1388 Ori, V380 Cyg and V413 Ser, for which the determination of the common age was possible but it was inconsistent with
the positions of the stars on the HR diagram.
In case of HI Mon a large overshooting $\alpha_{\rm ov}>0.5$ for the primary is necessary to get a common range of the age but then the primary is underluminous.
To agree the age of the components of the IM Mon system, it was necessary to adopt the higher metallicity, $Z=0.025$. However, with such high value of $Z$,
the secondary star of IM Mon is much overluminous.
In the case of V1388 Ori, establishing a common age required the large overshooting parameter $\alpha_{\rm ov}>0.5$ for the primary.
However, in the HR diagram both components are much overluminous.
More observations and analysis {\bf are} needed for these systems to reveal their actual state and formation history.
The binary V413 Ser is a completely different case, because most probably both components of this system are pre-main sequence stars
and here we did not considered this phase of evolution.

In the case of V380 Cyg, evolutionary computations with no-overshooting and metallicty $Z=0.014$, locate the primary beyond the main sequence
and no common age is determinable.  If one wants to keep the primary star on MS or just after it, a large overshooting, $\alpha_{\rm ov}=0.5$,
has to be included.
We cannot prove if such large core overshooting has a physical meaning.
From asteroseismic studies usually lower values, $\alpha<0.4$, are derived \citep[e.g.,][]{Pamyatnykh2004, Aerts2003, JDD2017}.
The important results is that usually for more massive components which rotate fast, we had to increase overshooting
to adjust the common age, e.g., V578 Mon, HI Mon, V380 Cyg, V1388 Ori.
It is possible that this large value of the overshooting parameter compensates for other mixing effect \citep{Jermyn2018}.
In particular it can be a signature of the rotational-induced mixing which we did not take into account.

To distinguish between various mixing processes, we need more observables which can be directly compared with theoretical counterparts.
It seems that most suitable systems for such studies are binaries with pulsating components.
The combined binary and asteroseismic modelling has the greatest potential to provide the most stringent constraints
on various parameters of a model and theory. This type of studies is still waiting for its golden age.

\acknowledgments
The work was financially supported by the Polish NCN grants 2015/17/B/ST9/02082 and 2018/29/B/ST9/02803.

\bibliography{JDDbiblio}
\bibliographystyle{aasjournal}

\appendix

\section{Stellar Binaries with the consistent age}

In Table\,3, we  give the results for 33 B-type binaries for which a consistent age exist, that is,
it was possible to derive a common age from the radius-age relation and to agree the star's positions on the HR diagram.
In Table\,4 we provide the whole range of the parameters ($Z,~\alpha_{\rm ov}$)
for which the consistent age was determinable.
The columns are self-explanatory in both tables to indicate the name and HD number (system), components, spectral types (SpT), masses ($M$), radii ($R$),
ages ($\log t$) of components, common age ($\log t_{AB}$), the mass fraction of hydrogen in the centre ($X_c$), metallicity, $Z$, and core overshooting ($\alpha_{ov}$). The ranges of $\alpha^A_{ov}$ and $\alpha^B_{ov}$  are given for $Z=0.014$ or for the nearest value of $Z$
for which the age was determinable.  For a given component, the overshooting parameter of the other component was fixed at $\alpha_{ov}=0.0$.
\startlongtable
\begin{deluxetable*}{lclcccccccrr}
	\tablecaption{The age of double-lined eclipsing B-type binaries for which consistent solutions of the age have been found.\label{chartable}}
	\tabletypesize{\scriptsize}
	\tablehead{
				\colhead{System} & \colhead{Star} &
		\colhead{SpT} & \colhead{$M$} &
		\colhead{$R$} & \colhead{$\log t$} &
		\colhead{$\log t_{AB}$} & \colhead{$X_c$} &
		\colhead{$Z$ } & \colhead{$\alpha_{\rm ov}$}\\
		\colhead{} & \colhead{} & \colhead{} & \colhead{[$M_{\odot}$]} &
		\colhead{[$R_{\odot}$]} & \colhead{[yr]} & \colhead{[yr]} &
		\colhead{} & \colhead{} & \colhead{}
	}
	\startdata
	\textbf{AH Cep}$^1$	& A	& B0.5Vn	& 15.26$\pm$0.35	& 6.346$\pm$0.071	&  $6.68\pm0.05$		& 6.69 -- 6.73		& 0.51 -- 0.44			& 0.014	& 0.0 \\
	HD 216014	          & B	& B0.5Vn	& 13.44$\pm$0.25	& 5.836$\pm$0.085	&  $6.75\pm0.06$		& 					& 0.52 -- 0.47			&		& 0.0 \vspace{5pt}\\
	\textbf{V578 Mon}	& A	& B1V		& 14.54$\pm$0.08		& 5.41$\pm$0.04		& $6.56\pm0.03$		& 6.59	 & 0.59 -- 0.57		& 0.011	& 0.3 \\
	HD 259135        	& B	& B2V		& 10.29$\pm$0.06		& 4.29$\pm$0.05		& $6.65\pm0.06$		& 				 & 0.61 -- 0.59		& 		& 0.0 \vspace{5pt}\\
	&	&			&	     				&					& $6.45\pm0.04$		& 6.46 -- 6.49	 & 0.61 -- 0.59		& 0.014	& 0.0 \\
	&	&			&					&					& $6.54\pm0.08 $ 	& 	 			 & 0.64 -- 0.62		& 		& 0.0 \vspace{5pt}\\
	\textbf{V453 Cyg }	& A	& B0.4IV		& 13.82$\pm$0.35		& 8.445$\pm$0.068			& $7.04\pm0.03$	    & 7.01 -- 7.06			& 0.24 -- 0.21		& 0.009	& 0.3 \\
	HD 227696	& B	& B0.7IV			& 10.64$\pm$0.22		& 5.420$\pm$0.068			& $7.05\pm0.04$		& 								& 0.39 -- 0.34					& 				& 0.0 \vspace{5pt}\\
	\textbf{V380 Cyg} $^2$ & A	& B1.5II-III & 13.13$\pm$0.24  & 16.22$\pm$0.26   & $7.20\pm0.02$  & 7.18 -- 7.20  & post--MS$^3$& 0.011 &  0.5 \\
	HD 187879	             & B	& B2V	  &  7.78$\pm$0.10  & 4.060$\pm$0.084  & $7.15\pm0.05$  & 	          & 0.52 -- 0.46 &	 & 0.0 \vspace{5pt}\\
	\textbf{CW Cep	}	& A	& B0.5V		& 13.05$\pm$0.20		& 5.64$\pm$0.12			& $6.76\pm0.06$		& 6.70 -- 6.82		& 0.53 -- 0.47					& 0.014	& 0.0 \\
	HD 218066	& B	& B0.5V			& 11.91$\pm$0.20		& 5.14$\pm$0.12			& $6.74\pm0.08$		& 					& 0.57 -- 0.51					& 				& 0.0 \vspace{5pt}\\
	\textbf{NY Cep}	  & A	& B0.5V			& 13.0$\pm$1.0 			& 6.8$\pm$0.7			& $6.94\pm0.15$ 	& 6.97 -- 7.09		& 0.45 -- 0.19					& 0.014	& 0.0 \\
	HD 217312	& B	& B2V 				& 9.0$\pm$1.0			& 5.4$\pm$0.5			& $7.18\pm0.21$		& 					& 0.48 -- 0.20					&				& 0.0 \vspace{5pt}\\
	\textbf{V346 Cen}		& A	& B1.5III		& 11.8$\pm$1.4  		& 8.2$\pm$0.3			& $7.15\pm0.11$ 	& 7.04 -- 7.25			& 0.30 -- 0.17				&  0.014	& 0.3   \\
	HD101837		& B	& B2V			& 8.4$\pm$0.8			& 4.2$\pm$0.2			& $6.92\pm0.34$ 	& 						& 0.64 -- 0.44 				&   		& 0.0     \vspace{5pt}\\
	\textbf{DW Car}	& A	& B1V			& 11.34$\pm$0.18			& 4.561$\pm$0.050		& $6.50\pm0.10$	& 6.40 -- 6.58		& 0.64 -- 0.60				& 0.012	& 0.0 \\
	HD 305543	& B	& B1V			& 10.63$\pm$0.20			& 4.299$\pm$0.058		& $6.43\pm0.15$ 	& 					& 0.66 -- 0.62			    & 		& 0.0 \vspace{5pt}\\
	\textbf{V379 Cep}	& A	& B2IV			& 10.56$\pm$0.23  		& 7.909$\pm$0.120		& $7.29\pm0.03$	& 7.26 -- 7.27 	& 0.10 -- 0.06			& 0.008 & 0.1 $^4$ \\
	HD 197770	  & B	& ---			&         6.09$\pm$0.13			& 3.040$\pm$0.040		& $7.16\pm0.12$	& 					& 0.64 -- 0.58			& 		& 0.0 \vspace{5pt}\\
	\textbf{QX Car 	}	& A 	& B2V 		& 9.25$\pm$0.12 	& 4.291$\pm$0.091 		& $6.84\pm0.09$      & 6.78 -- 6.93		& 0.63 -- 0.58 			& 0.014	& 0.0 \\
	HD 86118 	& B 	& B2V 			& 8.46$\pm$0.12 	& 4.053$\pm$0.091 		& $6.89\pm0.11$  	    & 					& 0.64 -- 0.58 			& 				& 0.0 \vspace{5pt}\\
	\textbf{V399 Vul}	& A 	& B3IV-V 		& 7.57$\pm$0.08 	& 5.820$\pm$0.030 		& $7.44\pm0.01$ 	    & 7.43 -- 7.44		& 0.14 -- 0.12		    & 0.010	& 0.0  \\
	HD 194495 	& B 	& B4V 			& 5.46$\pm$0.03 	& 3.140$\pm$0.080 		& $7.38\pm0.06$      & 					& 0.56 -- 0.50			& 		& 0.0  \vspace{5pt}\\
	\textbf{V497 Cep} 	& A 	& B3V 			& 6.89$\pm$0.46 			& 3.69$\pm$0.03 		& $7.11\pm0.18$   	& 6.93 -- 7.29			& 0.60 -- 0.51		& 0.014	& 0.0 \\
	BD+61 2213	& B 	& B4V 				& 5.39$\pm$0.40 			& 2.92$\pm$0.03		& $6.90\pm0.43$   	& 						& 0.68 -- 0.59 		& 		& 0.0 \vspace{5pt}\\
	\textbf{V539 Ara }	& A 	& B3V 		 	& 6.240$\pm$0.066 			& 4.516$\pm$0.084 		& $7.56\pm0.03$    	& 7.53 -- 7.58			& 0.32 -- 0.28 		& 0.014	& 0.0  \\
	HD 161783 	& B 	& B4V 				& 5.314$\pm$0.060 			& 3.428$\pm$0.083 		& $7.53\pm0.05$ 		& 						& 0.50 -- 0.44		& 	    & 0.0 \vspace{5pt}\\
	\textbf{CV Vel} 		& A 	& B2.5V 		 	& 6.086$\pm$0.044 			& 4.089$\pm$0.036 		& $7.52\pm0.02$ 		& 7.50 -- 7.53			& 0.39 -- 0.36		& 0.014	& 0.0  \\
	HD 77464 	& B 	& B2.5V 			        & 5.982$\pm$0.035 			& 3.950$\pm$0.036 		& $7.51\pm0.02$ 		& 						& 0.41 -- 0.39		&		& 0.0 \vspace{5pt}\\
	\textbf{LT CMa	}	& A 	& B4V 			& 5.59$\pm$0.20 			& 3.59$\pm$0.07 		& $7.52\pm0.09$		& 7.43 -- 7.46			& 0.48 -- 0.40		& 0.012	& 0.0 \\
	HD 53303 	& B 	& B6.5V 			        & 3.36$\pm$0.14 			& 2.04$\pm$0.05 		&     $< 7.46$		& 						& 0.70 -- 0.65		&		& 0.0 \vspace{5pt}\\
	\textbf{AG Per} 		& A 	& B3.4V 	& 5.359$\pm$0.160 		& 2.995$\pm$0.071 		& $7.15\pm0.20$ 	    & 6.94 -- 7.08		& 0.65 -- 0.58		    & 0.014	& 0.0 \\
	HD 25833 	        & B 	& B3.5V 	& 4.890$\pm$0.130 		& 2.605$\pm$0.070 		&      $< 7.08$	    & 					& 0.70 -- 0.65			& 		& 0.0 \vspace{5pt}\\
	\textbf{U Oph} 		& A 	& B5V 		& 5.273$\pm$0.091 		& 3.484$\pm$0.021 		& $7.57\pm0.03$ 	    & 7.54 -- 7.60		& 0.53 -- 0.51			& 0.014	& 0.0 \\
	HD 156247 	        & B 	& B6V 		& 4.739$\pm$0.072 		& 3.110$\pm$0.034 		& $7.59\pm0.05$      & 					& 0.59 -- 0.55			& 		& 0.0 \vspace{5pt}\\
	\textbf{DI Her}	     & A 	& B5V 	 	& 5.170$\pm$0.110 		& 2.681$\pm$0.046 	        &      $< 6.44$		& 0.00 -- 6.44		&  0.70 -- 0.69		    & 0.018	& 0.0 \\
	HD 175227 	    & B 	& B5V 		& 4.524$\pm$0.066 	    & 2.478$\pm$0.046 		&     $< 6.45$		& 					& 0.70 -- 0.69			& 		& 0.0 \vspace{5pt}\\
	\textbf{EP Cru}	    & A 	& B5V 		& 5.020$\pm$0.130 		& 3.590$\pm$0.035 		& $7.67\pm0.05$		& 7.69 -- 7.72		& 0.44 -- 0.39			& 0.014	& 0.0 \\
	HD 109724	    & B 	& B5V	 	& 4.830$\pm$0.130 		& 3.495$\pm$0.034 		& $7.73\pm0.04$   	& 					& 0.43 -- 0.40			& 		& 0.0 \vspace{5pt}\\
	\textbf{V760 Sco }	& A 	& B4V 	 	& 4.969$\pm$0.090 		& 3.015$\pm$0.066   	& $7.41\pm0.11$		& 7.30 -- 7.32		& 0.60 -- 0.53			& 0.014	& 0.0 \\
	HD 147683 	   & B 	& B4V 		& 4.609$\pm$0.073 		& 2.641$\pm$0.066   	& $7.07\pm0.25$		& 					& 0.68 -- 0.62			& 		& 0.0 \vspace{5pt}\\
	\textbf{MU Cas 	}	& A 	& B5V 		& 4.657$\pm$0.093 		& 4.195$\pm$0.058 		& $7.89\pm0.03$		& 7.86 -- 7.88		& 0.24 -- 0.19			 & 0.014	& 0.0 \\
	BD+59 22 	   & B 	& B5V 		& 4.575$\pm$0.088 		& 3.670$\pm$0.057 		& $7.85\pm0.03$          & 					& 0.35 -- 0.30			& 		& 0.0 \vspace{5pt}\\
	\textbf{GG Lup} 		& A 	& B7V 	& 4.106$\pm$0.044 		& 2.380$\pm$0.025 		& $6.89\pm0.21$ 	 	& 6.67 -- 7.10		& 0.69 -- 0.66			& 0.014	& 0.0 \\
	HD 135876 	& B 	& B9V 			& 2.504$\pm$0.023 		& 1.726$\pm$0.019 		&   $< 7.15$		& 					& 0.70 -- 0.69			& 		& 0.0 \vspace{5pt}\\
	\textbf{V615 Per} 	& A 	& B7V 		& 4.075$\pm$0.055 		& 2.291$\pm$0.141 		& $6.70\pm0.80$ 	        & 5.89 -- 7.50 				& 0.70 -- 0.60				& 0.009	& 0.0 \\
	                                  & B 	& ---			& 3.179$\pm$0.051 	        & 1.903$\pm$0.094 		&      $< 7.54$ 	   & 							& 0.70 -- 0.64				& 		& 0.0 \vspace{5pt}\\
	\textbf{BD+03 3821}  & A 	& B8V 		& 4.040$\pm$0.110 		& 3.770$\pm$0.030 		&  $8.02\pm0.04$ 	& 7.98 -- 8.06			& 0.26 -- 0.22	& 0.013	& 0.0  \\
	HD 174884	& B 	& ---				& 2.720$\pm$0.110 		& 2.040$\pm$0.020 		&  $7.95\pm0.16$ 	& 						& 0.62 -- 0.56		& 		& 0.0 \vspace{5pt}\\
	\textbf{V1665 Aql}	& A 	& B9V 		& 3.970$\pm$0.400 		& 4.130$\pm$0.100 		& $8.12\pm0.14$		& 7.98 -- 8.06			& 0.26 -- 0.13			& 0.019	& 0.0  \\
	HD 175677        	& B 	& ---			& 3.660$\pm$0.370    	& 2.600$\pm$0.060 		& $7.71\pm0.35$ 		& 						& 0.64 -- 0.49			& 		& 0.0 \vspace{5pt}\\
	\textbf{$\zeta$ Phe}	& A 	& B6V 		& 3.921$\pm$0.045 		& 2.852$\pm$0.015 		& $7.84\pm0.03$ 	         & 7.81 -- 7.87			& 0.48 -- 0.45			& 0.012	& 0.0 \\
	HD 6882 	   & B 	& B8V 			& 2.545$\pm$0.026 		& 1.854$\pm$0.011 		& $7.80\pm0.08$  	& 						& 0.65 -- 0.63			& 		& 0.0 \vspace{5pt}\\
	\textbf{YY Sgr	}	& A 	& B5V 		& 3.900$\pm$0.130 		& 2.560$\pm$0.030 		& $7.60\pm0.13$  	& 7.46 -- 7.73			& 0.61 -- 0.55		& 0.014	& 0.0  \\
	HD 173140 	& B 	& B6V			& 3.480$\pm$0.090 		& 2.330$\pm$0.050 		& $7.61\pm0.18$  	&						& 0.64 -- 0.57		&		& 0.0 \vspace{5pt}\\
	\textbf{V398 Lac}	& A 	& B9V 		& 3.830$\pm$0.350 		& 4.890$\pm$0.180 		& $8.16\pm0.05$  	& 8.11 -- 8.21				& 0.19 -- 0.08			& 0.017	& 0.2  \\
	HD 210180	& B 	& ---			& 3.290$\pm$0.320 			& 2.450$\pm$0.110 		& $7.82\pm0.39$ 	   & 							& 0.64 -- 0.45			& 		& 0.0 \vspace{5pt}\\
	\textbf{$\chi^2$ Hya}    &  A & B8V 	& 3.605$\pm$0.078 		& 4.390$\pm$0.039 			& $8.24\pm0.03$  	& 8.21 -- 8.25		& 0.13 -- 0.10			& 0.014	& 0.1 \\
	HD 96314 	& B & B8V 		& 2.632$\pm$0.049 		& 2.159$\pm$0.030 			& $8.18\pm0.07$	        & 	    	                 & 0.55 -- 0.50 			& 		& 0.0 \vspace{5pt}\\
	\textbf{V906 Sco}   & A 	& B9V         	& 3.378$\pm$0.071 	           & 4.521$\pm$0.035 	& $8.33\pm0.03$  	&  8.30 -- 8.32            & 0.07 -- 0.00		& 0.014	& 0.1 \\
	HD 162724 	& B 	& B9V 	       & 3.253$\pm$0.069 	                   & 3.515$\pm$0.039 	& $8.29\pm0.03$  	& 			& 0.22 -- 0.18 		& 		& 0.0 \vspace{5pt}\\
	\textbf{$\eta$ Mus}	& A 	& B8V 		& 3.300$\pm$0.040 		& 2.140$\pm$0.020 		& $7.04\pm0.28$	& 6.76 -- 7.38				& 0.69 -- 0.67				& 0.018	& 0.0  \\
	HD 114911	& B 	& B8V		& 3.290$\pm$0.040 			& 2.130$\pm$0.040 		& $6.75\pm0.63$	        & 						& 0.70 -- 0.66				& 		& 0.0 \vspace{5pt}\\
	\textbf{V799 Cas}	& A 	& B8V 		& 3.080$\pm$0.400 		& 3.230$\pm$0.140 		& $8.34\pm0.20$		& 8.17 -- 8.54			& 0.34 -- 0.13			& 0.014	& 0.0  \\
	HD 18915 	& B 	& B8.5V			& 2.970$\pm$0.400 		& 3.200$\pm$0.140 		& $8.38\pm	0.21$   	& 						& 0.33 -- 0.11			& 		& 0.0 \vspace{5pt}\\
	\textbf{PV Cas 	}	& A 	& B9.5V 		& 2.816$\pm$0.050 		& 2.301$\pm$0.020 		& $8.09\pm0.08$	& 8.01 -- 8.16			& 0.59 -- 0.55			& 0.023 & 0.0 \\
	HD 240208 	& B 	& B9.5V 		& 2.757$\pm$0.054 			& 2.257$\pm$0.019 		& $8.09\pm0.09$ 	& 							& 0.59 -- 0.55			& 		& 0.0 \vspace{5pt}\\
	\enddata
	\tablecomments{$^1$\cite{Torres2010}, $^2$ \cite{Pavlovski2009}, $^3$ this is the post-Main sequence phase after the overall contraction, $^4$ To agree the common age and the star's positions on the HR diagram, we had to increase the initial hydrogen abundance up to $X_0=0.74$ 
}
\end{deluxetable*}

\startlongtable
\begin{deluxetable*}{lclcccccccrr}
	\tablecaption{ The range of the metallicity $Z$ and the overshooting parameter $\alpha_{\rm ov}$ of each component for which a consistent age exists. 
\label{chartable}\\}
	\tabletypesize{\scriptsize}
	\tablehead{
		\colhead{System} & \colhead{Star} &
		\colhead{$M$} & \colhead{$R$} &
		\colhead{$Z$ } & \colhead{$\alpha^A_{\rm ov}$} & \colhead{$\alpha^B_{\rm ov}$}
	}
	\startdata
	\textbf{AH Cep}$^1$	& A	& 15.26$\pm$0.35	& 6.346$\pm$0.071	& 0.005 -- 0.017	& 0.0 -- 0.5	& 0.0 -- 0.3 \\
HD 216014	& B	& 13.44$\pm$0.25	& 5.836$\pm$0.085	& 	& 	&  \vspace{5pt}\\
\textbf{V578 Mon}	& A	& 14.54$\pm$0.08	& 5.41$\pm$0.04			& 0.011 -- 0.016	& 0.0 -- 0.5	& 0.0 --  0.2 \\ 
HD 259135	& B	& 10.29$\pm$0.06	& 4.29$\pm$0.05			& 	& 	&  \vspace{5pt}\\
\textbf{V453 Cyg }	& A	& 13.82$\pm$0.35	& 8.445$\pm$0.068		& 0.005 -- 0.009	& 0.3 -- 0.5	& 0.0 -- 0.2 \\
HD 227696	& B	& 10.64$\pm$0.22	& 5.420$\pm$0.068		& 	& 	&  \vspace{5pt}\\
\textbf{V380 Cyg}  $^2$& A	& 13.13$\pm$0.24  & 16.22$\pm$0.26  	 & 0.005 -- 0.011	& 0.5	& 0.0 -- 0.1  \\
HD 187879	    & B	&   7.78$\pm$0.10  & 4.060$\pm$0.084  		& 	& 	&   \vspace{5pt}\\
\textbf{CW Cep}		& A	& 13.05$\pm$0.20		& 5.64$\pm$0.12			& 0.005 -- 0.016	& 0.0 -- 0.5	& 0.0 -- 0.5  \\
HD 218066	& B	& 11.91$\pm$0.20		& 5.14$\pm$0.12			& 	& 	&   \vspace{5pt}\\
\textbf{NY Cep}			& A	& 13.0$\pm$1.0 			& 6.8$\pm$0.7			& 0.005 -- 0.018	& 0.0 -- 0.5	& 0.0 -- 0.5  \\
HD 217312	& B	& 9.0$\pm$1.0			& 5.4$\pm$0.5			& 	& 	&   \vspace{5pt}\\
\textbf{V346 Cen}		& A	& 11.8$\pm$1.4  		& 8.2$\pm$0.3			& 0.005 -- 0.014	& 0.3 -- 0.5	& 0.0 -- 0.5   \\
HD101837		& B	& 8.4$\pm$0.8			& 4.2$\pm$0.2			& 	& 	&   \vspace{5pt}\\
\textbf{DW Car 	}	& A	& 11.34$\pm$0.18			& 4.561$\pm$0.050		& 0.005 -- 0.012	& 0.0 -- 0.5	& 0.0 -- 0.5  \\
HD 305543	& B	& 10.63$\pm$0.20		& 4.299$\pm$0.058		& 	& 	&   \vspace{5pt}\\
\textbf{V379 Cep}	& A	& 10.56$\pm$0.23  		& 7.909$\pm$0.120		&  0.006 -- 0.008	& 0.1	& 0.0 -- 0.5  \\
HD 197770	& B	& 6.09$\pm$0.13			& 3.040$\pm$0.040		& 	& 	&   \vspace{5pt}\\
\textbf{QX Car 	}	& A & 9.25$\pm$0.12 	& 4.291$\pm$0.091 		& 0.010 -- 0.019	& 0.0 -- 0.5	& 0.0 -- 0.5  \\
HD 86118 	& B & 8.46$\pm$0.12 	& 4.053$\pm$0.091 		& 	& 	&   \vspace{5pt}\\
\textbf{V399 Vul}	& A & 7.57$\pm$0.08 	& 5.820$\pm$0.030 		& 0.010	& 0.0	& 0.0 -- 0.5   \\
HD 194495 	& B & 5.46$\pm$0.03 	& 3.140$\pm$0.080 		& 	& 	&    \vspace{5pt}\\
\textbf{V497 Cep }	& A & 6.89$\pm$0.46 	& 3.69$\pm$0.03 		& 0.007 -- 0.030	& 0.0 -- 0.5	& 0.0 -- 0.5  \\
BD+61 2213	& B & 5.39$\pm$0.40 	& 2.92$\pm$0.03		& 	& 	&   \vspace{5pt}\\
\textbf{V539 Ara }	& A & 6.240$\pm$0.066 	& 4.516$\pm$0.084 		& 0.010 -- 0.017	& 0.0 -- 0.4	& 0.0 -- 0.5   \\
HD 161783 	& B & 5.314$\pm$0.060 	& 3.428$\pm$0.083 		& 	& 	&   \vspace{5pt}\\
\textbf{CV Vel} 		& A & 6.086$\pm$0.044 		& 4.089$\pm$0.036 		& 0.011 -- 0.018	& 0.0 -- 0.2	& 0.0 -- 0.2  \\
HD 77464 	& B & 5.982$\pm$0.035 	& 3.950$\pm$0.036 		& 	& 	&   \vspace{5pt}\\
\textbf{LT CMa	}	& A & 5.59$\pm$0.20 		& 3.59$\pm$0.07 		& 0.010 -- 0.012	& 0.0 -- 0.1	& 0.0 -- 0.5  \\
HD 53303 	& B & 3.36$\pm$0.14 		& 2.04$\pm$0.05 		& 	& 	&   \vspace{5pt}\\
\textbf{AG Per}		& A & 5.359$\pm$0.160 		& 2.995$\pm$0.071 		& 0.006 -- 0.020	& 0.0 -- 0.5	& 0.0 -- 0.5   \\
HD 25833 	& B & 4.890$\pm$0.130 		& 2.605$\pm$0.070 		& 	& 	&   \vspace{5pt}\\
\textbf{U Oph} 		& A & 5.273$\pm$0.091 		& 3.484$\pm$0.021 		& 0.014 -- 0.021	& 0.0 -- 0.5	& 0.0 -- 0.4  \\
HD 156247 	& B & 4.739$\pm$0.072 		& 3.110$\pm$0.034 		& 	& 	&   \vspace{5pt}\\
\textbf{DI Her 	}	& A & 5.170$\pm$0.110 		& 2.681$\pm$0.046 	    & 0.018 -- 0.029	& 0.0 -- 0.5 & 0.0 -- 0.5  \\
HD 175227 	& B & 4.524$\pm$0.066 	    & 2.478$\pm$0.046 		& 	& 	&   \vspace{5pt}\\
\textbf{EP Cru	}	& A & 5.020$\pm$0.130 		& 3.590$\pm$0.035 		& 0.011 -- 0.021	& 0.0 -- 0.5	& 0.0 -- 0.2 \\
HD 109724	& B & 4.830$\pm$0.130 		& 3.495$\pm$0.034 		& 	& 	&   \vspace{5pt}\\
\textbf{V760 Sco }	& A & 4.969$\pm$0.090 		& 3.015$\pm$0.066   	& 0.010 -- 0.016	& 0.0 -- 0.1	& 0.0 -- 0.5  \\
HD 147683 	& B & 4.609$\pm$0.073 		& 2.641$\pm$0.066   	& 	& 	&  \vspace{5pt}\\
\textbf{MU Cas }	& A & 4.657$\pm$0.093 		& 4.195$\pm$0.058 		& 0.006 -- 0.020	& 0.0 -- 0.1	& 0.0 -- 0.5  \\
BD+59 22 	& B & 4.575$\pm$0.088 		& 3.670$\pm$0.057 		& 	& 	&   \vspace{5pt}\\
\textbf{GG Lup} 		& A & 4.106$\pm$0.044 		& 2.380$\pm$0.025 		& 0.014 -- 0.015	& 0.0 -- 0.5	& 0.0 -- 0.5  \\
HD 135876 	& B & 2.504$\pm$0.023 		& 1.726$\pm$0.019 		& 	& 	&   \vspace{5pt}\\
\textbf{V615 Per }	& A & 4.075$\pm$0.055 		& 2.291$\pm$0.141 		& 0.009 -- 0.017	& 0.0 -- 0.5	& 0.0 -- 0.5   \\
& B & 3.179$\pm$0.051 		& 1.903$\pm$0.094 		& 	& 	&   \vspace{5pt}\\
\textbf{BD+03 3821}  & A & 4.040$\pm$0.110 	& 3.770$\pm$0.030 		& 0.006 -- 0.019	& 0.0 -- 0.5	& 0.0 -- 0.5   \\
HD 174884	& B & 2.720$\pm$0.110 	& 2.040$\pm$0.020 		& 	& 	&   \vspace{5pt}\\
\textbf{V1665 Aql}	& A & 3.970$\pm$0.400 	& 4.130$\pm$0.100 			& 0.019 -- 0.027	& 0.0 -- 0.5	&  0.0  \\
HD 175677	& B & 3.660$\pm$0.370 	& 2.600$\pm$0.060 			& 	& 	&   \vspace{5pt}\\
\textbf{$\zeta$ Phe}	& A & 3.921$\pm$0.045 	& 2.852$\pm$0.015 			& 0.010 -- 0.012	& 0.0 -- 0.4	& 0.0 -- 0.5  \\
HD 6882 	& B & 2.545$\pm$0.026 	& 1.854$\pm$0.011 			& 	& 	&   \vspace{5pt}\\
\textbf{YY Sgr	}	& A & 3.900$\pm$0.130 		& 2.560$\pm$0.030 			& 0.006 -- 0.017	& 0.0 -- 0.5	& 0.0 -- 0.5   \\
HD 173140 	& B & 3.480$\pm$0.090 	& 2.330$\pm$0.050 			& 	& 	&   \vspace{5pt}\\
\textbf{V398 Lac}	& A & 3.830$\pm$0.350 	& 4.890$\pm$0.180 			& 0.017 -- 0.030	& 0.2 -- 0.5	& 0.0    \\
HD 210180	& B & 3.290$\pm$0.320 	& 2.450$\pm$0.110 			& 	& 	&  \vspace{5pt}\\
\textbf{$\chi^2$ Hya}    &  A & 3.605$\pm$0.078 	& 4.390$\pm$0.039 			& 0.010 -- 0.018	& 0.1 -- 0.4	&  0.0 -- 0.5   \\
HD 96314 	& B & 2.632$\pm$0.049 		& 2.159$\pm$0.030 			& 	& 	&   \vspace{5pt}\\
\textbf{V906 Sco }	& A	& 3.378$\pm$0.071 	     & 4.521$\pm$0.035 			& 0.012 -- 0.018	& 0.1 -- 0.2	& 0.0 -- 0.2  \\
HD 162724 	& B  & 3.253$\pm$0.069 	    & 3.515$\pm$0.039 			& 	& 	&   \vspace{5pt}\\
\textbf{$\eta$ Mus}	& A & 3.300$\pm$0.040 			& 2.140$\pm$0.020 		& 0.018 -- 0.022	& 0.0 -- 0.5	& 0.0 -- 0.5   \\
HD 114911	& B & 3.290$\pm$0.040 			& 2.130$\pm$0.040 		& 	& 	&   \vspace{5pt}\\
\textbf{V799 Cas}	& A & 3.080$\pm$0.400 			& 3.230$\pm$0.140 			& 0.005 -- 0.024	& 0.0 -- 0.5	& 0.0 -- 0.5   \\
HD 18915 	& B & 2.970$\pm$0.400 			& 3.200$\pm$0.140 			& 	& 	&   \vspace{5pt}\\
\textbf{PV Cas 	}	& A & 2.816$\pm$0.050 			& 2.301$\pm$0.020 			& 0.023 -- 0.030	& 0.0 -- 0.2	& 0.0 -- 0.5  \\
HD 240208 	& B & 2.757$\pm$0.054 			& 2.257$\pm$0.019 		& 	& 	&   \vspace{5pt}\\
	\enddata
	\tablecomments{$^1$\cite{Torres2010}, 
		$^2$ \cite{Pavlovski2009}}
\end{deluxetable*}

\section{Detailed results of the age determination from the radius-age relation}
As mentioned in the main part of the paper, in the first step we determine the age from the mass-radius-age diagrams.
Here, we present the details of these determination for a sample binary, V578 Mon, and the results for all 38 binaries are available on-line at \url{http://www.astro.uni.wroc.pl/ludzie/miszuda/age_Btype.html}. Each file contains the values of the age for various combinations of the metallicity $Z$ and
the values of the overshooting parameter for the component A and B: $\alpha^A_{ov}$ and $\alpha^B_{ov}$.
We considered the following ranges: $Z = 0.005 - 0.030$ with a step 0.001 and $\alpha^{A,B}_{ov} = 0.0 - 0.5$ with a step 0.1.

Each file starts with a header in which the information about the mass and radii range used for age determination is given:

\begin{verbatim}
# (c) Amadeusz Miszuda, miszuda(at)astro.uni.wr
#
# Age determination done for set of parameters:
# M1max =  14.62  M1min =  14.46
# R1max =  5.450  R1min =  5.370
#
# M2max =  10.35  M2min =  10.23
# R2max =  4.340  R2min =  4.240
\end{verbatim}

After a header, there is a set of tables for different metallicities $Z$ with a step of $\Delta Z = 0.001$.
Each table has a fixed style; rows correspond to overshooting for the primary and columns - overshooting for the secondary. For each pair of $\alpha_{ov}$, the age with the error is given.
If there is no common age then  the zero values are inserted. Here, we appended a few values of $Z$ and a full set is available on-line, at \url{http://www.astro.uni.wroc.pl/ludzie/miszuda/age_Btype.html}. Note, that on-line material is given for a complete $Z$ range, not for those solutions only, which are consistent in the HR diagram.


\begin{widetext}
\begin{verbatim}
Z =0.005
ov(A)\ov(B)| 0.0         | 0.1         | 0.2         | 0.3         | 0.4         | 0.5
----------------------------------------------------------------------------------------------
       0.0 | 0.000 0.000 | 0.000 0.000 | 0.000 0.000 | 0.000 0.000 | 0.000 0.000 | 0.000 0.000
       0.1 | 0.000 0.000 | 0.000 0.000 | 0.000 0.000 | 0.000 0.000 | 0.000 0.000 | 0.000 0.000
       0.2 | 0.000 0.000 | 0.000 0.000 | 0.000 0.000 | 0.000 0.000 | 0.000 0.000 | 0.000 0.000
       0.3 | 0.000 0.000 | 0.000 0.000 | 0.000 0.000 | 0.000 0.000 | 0.000 0.000 | 0.000 0.000
       0.4 | 0.000 0.000 | 0.000 0.000 | 0.000 0.000 | 0.000 0.000 | 0.000 0.000 | 0.000 0.000
       0.5 | 0.000 0.000 | 0.000 0.000 | 0.000 0.000 | 0.000 0.000 | 0.000 0.000 | 0.000 0.000
\end{verbatim}

\begin{verbatim}
Z =0.010
ov(A)\ov(B)| 0.0          | 0.1         | 0.2         | 0.3         | 0.4         | 0.5
-----------------------------------------------------------------------------------------------
       0.0 |  0.000 0.000 | 0.000 0.000 | 0.000 0.000 | 0.000 0.000 | 0.000 0.000 | 0.000 0.000
       0.1 |  0.000 0.000 | 0.000 0.000 | 0.000 0.000 | 0.000 0.000 | 0.000 0.000 | 0.000 0.000
       0.2 |  0.000 0.000 | 0.000 0.000 | 0.000 0.000 | 0.000 0.000 | 0.000 0.000 | 0.000 0.000
       0.3 |  0.000 0.000 | 0.000 0.000 | 0.000 0.000 | 0.000 0.000 | 0.000 0.000 | 0.000 0.000
       0.4 |  0.000 0.000 | 0.000 0.000 | 0.000 0.000 | 0.000 0.000 | 0.000 0.000 | 0.000 0.000
       0.5 |  6.635 0.004 | 0.000 0.000 | 0.000 0.000 | 0.000 0.000 | 0.000 0.000 | 0.000 0.000
\end{verbatim}

\begin{verbatim}
Z = .0150
ov(A)\ov(B)| 0.0          | 0.1         | 0.2         | 0.3         | 0.4         | 0.5
-----------------------------------------------------------------------------------------------
       0.0 |  6.447 0.022 | 6.450 0.020 | 6.459 0.011 | 6.468 0.002 | 0.000 0.000 | 0.000 0.000
       0.1 |  6.450 0.025 | 6.453 0.023 | 6.462 0.014 | 6.471 0.005 | 0.000 0.000 | 0.000 0.000
       0.2 |  6.456 0.032 | 6.459 0.029 | 6.468 0.020 | 6.477 0.011 | 6.485 0.003 | 0.000 0.000
       0.3 |  6.463 0.038 | 6.466 0.035 | 6.474 0.027 | 6.483 0.018 | 6.492 0.009 | 6.500 0.001
       0.4 |  6.471 0.042 | 6.472 0.041 | 6.480 0.033 | 6.489 0.024 | 6.498 0.015 | 6.506 0.007
       0.5 |  6.484 0.042 | 6.484 0.042 | 6.487 0.039 | 6.496 0.030 | 6.504 0.022 | 6.513 0.014
\end{verbatim}

\begin{verbatim}
Z = .0200
ov(A)\ov(B)| 0.0          | 0.1         | 0.2         | 0.3         | 0.4         | 0.5
-----------------------------------------------------------------------------------------------
       0.0 |  6.297 0.057 | 6.297 0.057 | 6.297 0.057 | 6.297 0.057 | 6.298 0.057 | 6.306 0.048
       0.1 |  6.300 0.058 | 6.300 0.058 | 6.300 0.058 | 6.300 0.058 | 6.300 0.058 | 6.308 0.050
       0.2 |  6.314 0.057 | 6.314 0.057 | 6.314 0.057 | 6.314 0.057 | 6.314 0.057 | 6.315 0.057
       0.3 |  6.327 0.057 | 6.327 0.057 | 6.327 0.057 | 6.327 0.057 | 6.327 0.057 | 6.327 0.057
       0.4 |  6.341 0.057 | 6.341 0.057 | 6.341 0.057 | 6.341 0.057 | 6.341 0.057 | 6.341 0.057
       0.5 |  6.356 0.057 | 6.356 0.057 | 6.356 0.057 | 6.356 0.057 | 6.356 0.057 | 6.356 0.057
\end{verbatim}

\begin{verbatim}
Z = .0250
ov(A)\ov(B)| 0.0          | 0.1         | 0.2         | 0.3         | 0.4         | 0.5
-----------------------------------------------------------------------------------------------
       0.0 |  6.150 0.078 | 6.150 0.078 | 6.150 0.078 | 6.150 0.078 | 6.150 0.078 | 6.150 0.078
       0.1 |  6.150 0.078 | 6.150 0.078 | 6.150 0.078 | 6.150 0.078 | 6.150 0.078 | 6.150 0.078
       0.2 |  6.162 0.075 | 6.164 0.077 | 6.165 0.078 | 6.165 0.078 | 6.165 0.078 | 6.165 0.078
       0.3 |  6.170 0.068 | 6.172 0.070 | 6.179 0.078 | 6.179 0.078 | 6.179 0.078 | 6.179 0.078
       0.4 |  6.177 0.061 | 6.179 0.062 | 6.189 0.072 | 6.194 0.077 | 6.194 0.077 | 6.194 0.077
       0.5 |  6.185 0.053 | 6.186 0.055 | 6.197 0.065 | 6.206 0.074 | 6.209 0.077 | 6.209 0.077
\end{verbatim}

\end{widetext}

\newpage

\section{A comparison with determinations in the literature}

\startlongtable
\begin{deluxetable*}{llll}
	\tablecaption{A comparison of the values of the age of B-type DDLEBs determined by us and the corresponding values found in the literature.\label{chartable}\\}
	\tabletypesize{\scriptsize}
	\tablehead{
		\colhead{Name} & \colhead{$\log t_{\rm our}$} & \colhead{$\log t_{\rm lit}$	} & \colhead{Ref.}}
	\startdata
AH Cep $^1$        			&	6.71 $\pm$ 0.02					&	6.68 $\pm$ 0.18						&	\cite{Holmgren1990} \\
V578 Mon						&	6.48 $\pm$ 0.01				&	6.36 $\pm$ 0.04	    					&	\cite{Hensberge2000} \\
V453 Cyg						&	7.02 $\pm$ 0.01				&	7.00 $\pm$ 0.08						&	\cite{Pavlovski2009A} \\
CW Cep							&	6.76 $\pm$ 0.06					&	7.00 $\pm$ 0.05 						&	\cite{Clausen1991} \\
DW Car							&	6.49 $\pm$ 0.09     			&	6.34 $\pm$ 0.26						&	\cite{Southworth2007} \\
QX Car							&	6.85 $\pm$ 0.07      			&	6.90 $\pm$ 0.13						&	\cite{Andersen1983} \\
V359 Ara						&	7.56 $\pm$ 0.03		     		&	7.60		    				&	\cite{Clausen1996}\\
CV Vel							&	7.52 $\pm$ 0.02					&	7.47							&	\cite{Clausen1977}\\
AG Per							&	7.01 $\pm$ 0.07			    	&	7.69 $\pm$ 0.09						&	\cite{Gimenez1994}\\
U Oph							&	7.57 $\pm$ 0.03					&	7.60							&	\cite{Vaz2007}\\
DI Her							&	     $<$ 6.44   				&	6.65 $\pm$ 0.35						&	\cite{Claret2010}\\
V760 Sco						&	7.31 $\pm$ 0.01					&	7.40 $\pm$ 0.20						&	\cite{Andersen1985}\\
MU Cas 	    					&	7.87 $\pm$ 0.01					&	7.81 $\pm$ 0.04							&	\cite{Lacy2004}\\
GG Lup							&	6.89 $\pm$ 0.22     			&	7.30							&	\cite{Andersen1993}\\
$\zeta$ Phe						&	7.84 $\pm$ 0.03					&	7.60							&	\cite{Torres2010}\\
$\chi^2$ Hya					&	8.23 $\pm$ 0.03					&	8.25 $\pm$ 0.05		        			&	\cite{Clausen1978}\\
V906 Sco  						&	8.31 $\pm$ 0.01								&	8.370 $\pm$ 0.28        				&	\cite{Sestito2003}\\
PV Cas							&	8.09 $\pm$ 0.07				&	8.30							&	\cite{Torres2010}\\
V615 Per						&	6.69 $\pm$ 0.81				&	7.10 $\pm$ 0.01	                 		&	\cite{Southworth2004A}\\
V497 Cep						&	7.11 $\pm$ 0.18				&	7.25 $\pm$ 0.11	        			&	\cite{Conti1970}\\
NY Cep							&	7.03 $\pm$ 0.06	    			&	6.77							&	\cite{Albrecht2011}\\
V380 Cyg       					&	7.19 $\pm$ 0.01					&	7.24 $\pm$ 0.39   					&	\cite{Claret2003}\\
V399 Vul						&	7.44 $\pm$ 0.01     			&	7.45 $\pm$ 0.05	         				&	\cite{Cakirli2012}\\
LT CMa							&	7.44 $\pm$ 0.01					&	7.54 $\pm$ 0.07			    			&	\cite{Bakis2010}\\
EP Cru							&	7.71 $\pm$ 0.02				    &	7.75 $\pm$ 0.04	        				&	\cite{Albrecht2013}\\
BD+03\_3821						&	8.02 $\pm$ 0.04					&	8.10 $\pm$ 0.25						&	\cite{Maceroni2009}\\
V1665 Aql						&	8.02 $\pm$ 0.04					&	8.20							&	\cite{Ibanoglu2007}\\
YY Sgr							&	7.60 $\pm$ 0.13				&	7.47							&	\cite{Lacy1997}\\
$\eta$ Mus						&	7.04 $\pm$ 0.28	    		&	7.20							&	\cite{Bakis2007}\\
V799 Cas						&	8.35 $\pm$ 0.19				&	8.39							&	\cite{Ibanoglu2009}\\
	\enddata
	\tablecomments{$^1$\cite{Torres2010}}
\end{deluxetable*}

\label{lastpage}
\end{document}